\documentclass[sigconf]{acmart} % This will produce a double-column layout
\usepackage{framed}
\AtBeginDocument{%
  }

\copyrightyear{2025} 
\acmYear{2025} 
\acmConference[UMAP '25]{33rd ACM Conference on User Modeling, Adaptation
and Personalization}{June 16--19, 2025}{New York City, NY, USA}
\acmBooktitle{33rd ACM Conference on User Modeling, Adaptation and
Personalization (UMAP '25), June 16--19, 2025, New York City, NY, USA}
\acmDOI{10.1145/3699682.3728343}
\acmISBN{979-8-4007-1313-2/2025/06}

\begin{document}

\title{``\textit{Even explanations will not help in trusting} [this] \textit{fundamentally biased system}'': A Predictive Policing Case-Study}

\author{Siddharth Mehrotra}
\affiliation{%
  \institution{Delft University of Technology}
  \streetaddress{Van Mourik Broekmanweg 6}
  \city{Delft}
  \country{The Netherlands}
  \postcode{2628 XE}
}\email{s.mehrotra@tudelft.nl}

\author{Ujwal Gadiraju}
\affiliation{%
  \institution{Delft University of Technology}
  \streetaddress{Van Mourik Broekmanweg 6}
  \city{Delft}
  \country{The Netherlands}
  \postcode{2628 XE}
}\email{u.k.gadiraju@tudelft.nl}

\author{Eva Bittner}
\affiliation{%
  \institution{University of Hamburg}
  \streetaddress{}
  \city{Hamburg}
  \country{Germany}
  \postcode{}
}\email{eva.bittner@uni-hamburg.de}

\author{Folkert van Delden}
\affiliation{%
  \institution{Delft University of Technology}
  \streetaddress{Van Mourik Broekmanweg 6}
  \city{Delft}
  \country{The Netherlands}
  \postcode{2628 XE}
}\email{f.e.vandelden@tudelft.nl}

\author{Catholijn M. Jonker}
\affiliation{%
  \institution{Delft University of Technology \& LIACS, Leiden University}
  \streetaddress{Van Mourik Broekmanweg 6}
  \city{Delft}
  \country{The Netherlands}
  \postcode{2628 XE}}
\email{c.m.jonker@tudelft.nl}

\author{Myrthe L. Tielman}
\affiliation{%
  \institution{Delft University of Technology}
  \streetaddress{Van Mourik Broekmanweg 6}
  \city{Delft}
  \country{The Netherlands}
  \postcode{2628 XE}
}\email{m.l.tielman@tudelft.nl}

\renewcommand{\shortauthors}{Mehrotra et al.}

\begin{abstract}
In today's society, where Artificial Intelligence (AI) has gained a vital role, concerns regarding user's trust have garnered significant attention. The use of AI systems in high-risk domains have often led users to either under-trust it, potentially causing inadequate reliance or over-trust it, resulting in over-compliance. Therefore, users must maintain an appropriate level of trust. Past research has indicated that explanations provided by AI systems can enhance user understanding of when to trust or not trust the system. However, the utility of presentation of different explanations forms still remains to be explored especially in high-risk domains. Therefore, this study explores the impact of different explanation types (text, visual, and hybrid) and user expertise (retired police officers and lay users) on establishing appropriate trust in AI-based predictive policing. While we observed that the hybrid form of explanations increased the subjective trust in AI for expert users, it did not led to better decision-making. Furthermore, no form of explanations helped build appropriate trust. The findings of our study emphasize the importance of re-evaluating the use of explanations to build [appropriate] trust in AI based systems especially when the system's use is questionable. Finally, we synthesize potential challenges and policy recommendations based on our results to design for appropriate trust in high-risk based AI-based systems.
\end{abstract}

\begin{CCSXML}
<ccs2012>
   <concept>
       <concept_id>10010147.10010178</concept_id>
       <concept_desc>Computing methodologies~Artificial intelligence</concept_desc>
       <concept_significance>500</concept_significance>
       </concept>
   <concept>
       <concept_id>10003120.10003121</concept_id>
       <concept_desc>Human-centered computing~Human computer interaction (HCI)</concept_desc>
       <concept_significance>500</concept_significance>
       </concept>
   <concept>
       <concept_id>10003456.10003462.10003588</concept_id>
       <concept_desc>Social and professional topics~Government technology policy</concept_desc>
       <concept_significance>500</concept_significance>
       </concept>
 </ccs2012>
\end{CCSXML}

\ccsdesc[500]{Computing methodologies~Artificial intelligence}
\ccsdesc[500]{Human-centered computing~Human computer interaction (HCI)}
%\ccsdesc[500]{Social and professional topics~Government technology policy}

\keywords{Appropriate Trust, Explainable AI, Predictive Policing, Trust}

\maketitle

\section{Introduction}
Artificial Intelligence (AI) is rapidly reshaping public organizations globally, mainly through machine learning approaches that automate routine administrative tasks and support decision-making \cite{bullock2019artificial}. One of the key components to achieving effective decision-making is a user's appropriate trust in AI systems. Despite recent efforts towards enhancing trust in algorithmic decision-making systems (e.g., adding \textit{explanations} \cite{dodge2019, yang2020visual}, \textit{human oversight} \cite{suzanne2022capable, ososky2013building, shively2018human}, and \textit{confidence scores} \cite{zhang2020effect, barbosa2022investigating}), comparatively little attention has been paid to building appropriate trust in them. Both under-trust and over-trust are deemed inappropriate \cite{parasuraman1997humans,robinette2016overtrust,leesee}, instead we require trust to be appropriate. Under-trust can lead to under-reliance, and over-trust can lead to over-compliance, which can negatively impact the task. Therefore, in this work we study whether we can improve the appropriateness of trust in an AI decision support system (\textit{goal}). We propose to do this through explanations (\textit{means}), and position our work in the context of AI-based predictive policing (\textit{context}) as a high-risk domain.

We choose AI-based predictive policing as our use case primarily because it represents a critical domain where appropriate trust is pivotal due to the high-stakes nature of decision-making. Additionally, the use of AI in predictive policing has been the subject of extensive debate for years, with numerous studies highlighting biases inherent in data collection practices and their impact on AI-driven decisions \cite{marda2020data,meijer2019predictive,meijer2021algorithmization,ferguson2018legal,alikhademi2022review,leese2024staying}. Despite these concerns, predictive policing systems continue to be used globally \cite{mugari2021predictive}. Therefore, in this study we explore whether providing explanations can help users critically evaluate AI decision-making in this use-case, encouraging introspection and ultimately NOT promoting trust but fostering appropriate trust.

We draw inspiration from the ASPECT model by Jameson et al. \cite{jameson2014choice}, as a useful means to identify the different facets of our complex use case. The ASPECT acronym refers to the first letters of six patterns: Attributes, Social influence, Policies, Experience, Consequences, and Trial and error. Each pattern may be perceived as an "aspect" of design choices, denoting a particular perspective or approach to addressing a problem or research question. Aligned with the ASPECT model, our approach with this use-case is based on attributes (explanations), social (socially relevant topic), policy (policy level implications), experience (user expertise), consequence (effect of explanation), and trial \& error (human-AI agreement) based choice patterns of the ASPECT model. %Furthermore, examining how different explanations impact appropriate trust in this context is practical as it informs the design and deployment of AI systems particularly in sensitive areas like law enforcement, promoting responsible and effective use \cite{meijer2019predictive}. %

%Explainable AI (XAI) is meant to give insight into the AI’s internal model and decision-making \cite{wang2022effects} and has been shown to help users understand how the system works \cite{cai2019effects,paez2019pragmatic}. Efforts to ensure that AI is trusted appropriately are often in the form of explanations \cite{mehrotraintegrity,bansal2021does,liu2021understanding}. While there are some prior works showing explanations not helping in trust calibration \cite{bansal2021does,zhang2020effect,zana}, several other recent works actually show the usefulness of explanations for this purpose \cite{10.1145/3579605,10.1145/3631614,10.1145/3610218,mehrotraintegrity}. Given this clear lack of consensus on the helpfulness of explanations for appropriate trust, this topic requires further granular exploration.

While AI systems can provide explanations in multiple formats, selecting the optimal design remains challenging. These formats include visual explanations like saliency maps \cite{adadi2018peeking}, textual explanations using words and phrases, and analytical explanations that enable users to explore both data and model \cite{kim2018textual,he2022like}. Although each method has garnered significant attention independently, comparative studies examining their relative effectiveness across different contexts and user groups remain limited \cite{szymanski2021visual, robbemond2022understanding, park2018multimodal, nunes2017systematic}. Notably, there is a particular gap in understanding how these different presentation methods impact appropriate trust in high-risk domains. Our research addresses this gap by examining user interaction and perception across textual, visual, and hybrid explanation formats to determine their effectiveness in fostering appropriate trust.

According to Ribera \& Lapedriza \cite{ribera2019can}, the goal of XAI in any context depends not only on the presentation of explanations but also on the type of end-user that is on the receiving end of the explanations. For example, previous research has shown that the hybrid form of explanations significantly improves correct understanding for lay users compared to visual explanations \cite{szymanski2021visual}. However, little work has been done so far to compare the utility of different explanation methods in building appropriate trust between expert users and lay users \cite{larasati2023meaningful}. In the predictive policing use-case, this comparison is relevant as some police officers might have less professional experience to draw on than others, for example, police officers who have recently joined the department. Therefore, in this study, we compared the utility of different types of explanations with both expert and lay users (\textit{moderation factor}) linked with the experience pattern of the ASPECT model. 

To better understand the moderating factor of user expertise in our study, we performed an application-grounded evaluation followed by a human-grounded evaluation. Human-grounded evaluation is appealing when orchestrating experiments with the target user groups is challenging \cite{10.1145/2939672.2939874,kim2015inferring}. This is in line with Doshi-Velez \& Kim \cite{doshi2017towards} who articulate and argue for the value of different layers of evaluation within XAI focusing on functionality, grounding, and presentation. Our sample of police officers corresponds to application-grounded evaluation which includes experts and real tasks. On the other hand, our sample of lay users corresponds to human-grounded evaluation which includes real humans performing identical tasks. 
%Furthermore, we echo Doshi-Velez \& Kim \cite{doshi2017towards} recommendation that ``\textit{Just as one would be critical of a reliability-oriented paper that only cites accuracy statistics, the choice of evaluation should match the specificity of the claim being made.}" 
Our contribution in this work is centered around a specific use-case that warrants assessment within the framework of application and human-grounded evaluation.

We aim to address the following research questions:

\noindent \textbf{RQ1:} What effect do different types of explanations (no explanation, textual, visual or hybrid) have on building appropriate trust in AI-based predictive policing systems?\\
 \textbf{RQ2:} How does human trust in the AI assistant change given these different types of explanations? \\
 \textbf{RQ3:} Do lay users or experts find these explanations useful in making a decision?\\
 \textbf{RQ4:} Is there a moderating effect of user expertise influence the role of explanations in establishing appropriate or subjective trust?

We investigated the first question by prompting users with a selection of hotspots for predictive policing that gauge their understanding of the explanation at hand and see whether they can appropriately trust the system. For measuring appropriate trust, we adopt several existing measures from the literature. The second question is addressed by asking participants to rate their perceived trust in the AI assistant, and the third by rating the usefulness of the explanations over multiple rounds in our study. Finally, we address the last question by comparing participants with different expertise (police officers \& lay users) and studying the role of explanations in building appropriate and subjective trust.

\textbf{Original Contributions.} Through our work in this paper, we make the following contributions:
%\vspace{-0.19cm}
\begin{enumerate}
    \item  We present the first study exploring the effect of different form of explanations on building appropriate trust using a prototypical system in
                the context of AI-based predictive policing.
    \item  We illustrate the effect of user expertise on different form of explanations for building appropriate trust.
    \item By conducting two user studies (\textit{N=192, 12 experts and 180 layusers}), we show that even with different form of explanations, participants often end up in the trap of confirmation biases and no form of explanation helped in fostering appropriate trust.
    \item Based on our results, we highlight research challenges and recommendations for the design of public sector AI systems.
\end{enumerate}    
\textbf{What this work is \underline{not} about?\footnote{The decision to exclude discussions on the ethical implications of predictive policing and the individuals affected by such systems is not one made lightly, but rather stems from the specific focus of our research. While acknowledging the profound ethical considerations surrounding predictive policing, our study is primarily concerned with investigating the role of different forms of explanations in building appropriate trust.}} 
\begin{enumerate}
    \item  Ethical implications of predictive policing and people affected by such systems.
    \item  Target users of the AI-based predictive policing system.
\end{enumerate}

%It's crucial to note that our research does not endorse or promote the application of predictive policing systems per se. Instead, we aim to provide insights into the factors that influence users' appropriate trust in such a use-case, recognizing its ethical significance within a broader societal context. Thus, while the ethical implications of predictive policing are undoubtedly pertinent and warrant thorough examination, our study adopts a narrower scope focused specifically on examining XAI forms for fostering appropriate trust.

\section{Background and Related Work}
\subsection{Appropriate Trust}
Appropriate Trust in AI systems has rapidly become an important area of focus for researchers and practitioners. As technology evolved from automated machines to decision aids, virtual avatars, robots, and AI teammates, appropriate trust has been studied in depth and breadth across various domains. %Appropriate trust is often linked to the alignment between the perceived and actual performance of the system \cite{yang2020visual}. Mehrotra et al. argue that human trust in the AI system must be appropriate because, with appropriate trust in AI, people may be simultaneously aware of AI's potential and limitations \cite{mehrotraintegrity}. This should reduce the harms and negative consequences of misuse and disuse of AI \cite{leesee}.

\subsubsection{Definitions}
It is important to understand how we define appropriate human trust in AI (Human-AI trust) when trying to achieve it. On the one hand, the definitions of appropriate trust are linked to system performance or reliability, such as by \citet{mcbride2010trust, mcguirl2006supporting,yang2020visual,ososky2013building}. On the other hand, the definitions of appropriate trust are related to trustworthiness and beliefs such as by \citet{carolina,danks2019value,mehrotraintegrity}.
Inevitably, despite the crucial role of appropriate trust in ensuring the successful use of AI systems, there is currently a fragmented overview of its understanding \cite{liao2022designing}. This conclusion resonates with \cite{jacovi} overview, who calls for definitions to be precisely defined and differentiated.

\subsubsection{Use of Explanations for Fostering Appropriate Trust}
A common method to achieve appropriate trust is by adding transparency to the system through explanations \cite{10.1145/3579605,10.1145/3631614,10.1145/3610218,mehrotraintegrity}. Intuitively, this makes sense as understanding an AI system’s inner workings and decision-making should, in theory, also allow a user to understand better when to trust or not trust a system to perform a task \cite{10.1145/3613905.3650825}. For example, it has been shown that an AI agent who displays its integrity in the form of explanations by being explicit about potential biases in data or algorithms achieved appropriate trust more often than being honest about capability or transparent about the decision-making process \cite{mehrotraintegrity}. Similarly, Yang et al. results indicated that visual explanation led to users’ appropriate trust in machine learning and improved appropriate use of the recommendations from the classifier \cite{yang2020visual}.

On the flip side of the coin, some empirical evidence suggests that even technically sound AI explanations can result in harmful over-trust and over-reliance \cite{bansal2021does,10.1145/3290607.3312787,kaur2020interpreting,zhang2020effect}. For example, studying about ‘cognitive forcing functions’ Buçinca et al. have shown that explanations with these functions were effective in trust calibration, as here the AI system adjusts to the user’s attitude and behaviour following the signs of over- and under-trust \cite{zana}. But the study, in contrast, highlights that people do not cognitively engage with explanations. Similarly, Bertrand et al. find that providing feature-based explanations does not improve appropriate reliance or understanding compared to not providing any explanation \cite{bertrand2023questioning}. Therefore, given this clear lack of consensus more work into the effect of XAI on appropriate trust is warranted \cite{liao2022designing}.

Other research in XAI has explored how expertise influences the perception of explanations in building trust. For instance, Simkute et al. \cite{simkute} highlight the importance of tailoring explanations to account for differences in reasoning between experts and lay users. Naturally, experts tend to critique explanations more rigorously, which can sometimes lead to insufficient trust, whereas lay users are more prone to over-reliance on the system \cite{bayer2022role,schaffer2019can}. A common denominator highlights that explanations must support either trust building for experts, or critical thinking for lay users.

\subsection{AI-based Predictive Policing \&  Trust}
In light of the EU AI Act's regulatory framework, the use of predictive policing systems, which heavily relies on advanced data analysis methods, may come under scrutiny and be subject to compliance with the Act's provisions on high-risk AI applications in law enforcement \cite{com2021laying}. This introduces a dimension of accountability and transparency in deploying predictive policing technologies within the legal and ethical parameters \cite{hardyns2018predictive, alikhademi2022review}. 

Currently, there are four main applications of predictive policing being used in European and American police departments: CAS (Crime Anticipation System) in the Netherlands\footnote{\url{https://kombijde.politie.nl/vakgebieden/ict/predictiv}},  ProMap and PredPol in UK\footnote{\url{https://www.lawsociety.org.uk/topics/research/algorithm-use-in-the-criminal-justice-system-report}}, and Soundthinking in the US\footnote{\url{https://www.soundthinking.com/law-enforcement/crime-analysis-crimetracer/}}. Multiple studies have been conducted to highlight the issues with these applications \cite{oosterloo2018politics,gerstner2018predictive,max2022soundthinking}. For example, Meijer et al.'s study highlights two patterns of algorithmization of government bureaucracy - the 'algorithmic cage’ (Berlin, more hierarchical control) and the ‘algorithmic colleague’ (Amsterdam, room for professional judgment) \cite{meijer2021algorithmization}. Specifically looking at trust, a study by Selten et al. shows that police officers trust and follow AI recommendations congruent with their intuitive professional judgment \cite{selten}. 

Studies examining predictive policing systems consistently reveal systemic limitations and biases, particularly regarding the over-policing of marginalized communities \cite{lum2016predict,ensign2018runaway,akpinar2021effect}. Lum \& Isaac's research demonstrates how these systems amplify historical biases, intensifying surveillance in already over-policed areas while simultaneously reducing algorithmic accountability \cite{lum2016predict}. This issue is compounded by the self-reinforcing feedback loops identified by Ensign et al., where police resources are repeatedly directed to the same neighborhoods independent of actual crime rates \cite{ensign2018runaway}. Overall, these findings underscore the critical importance of appropriate trust in predictive policing systems, making this domain an essential case study for examining how explanations can foster appropriate trust. %Building upon these prior works and utilizing predictive policing as a real-world use-case, we explore the role of different explanation forms and user expertise on appropriate trust in this domain.

\section{Study Design}
In our main user study, we sought to understand the effect of different explanations on appropriate trust and the role of user expertise. To understand the role of user expertise, we conducted two user studies. In the first user study, we recruited 12 ex-police officers from the Dutch police who retired in the last five years. We refer to this group of participants as "\textit{Expert users}". In the second user study, we recruited 180 crowdsourced workers without experience with predictive policing systems. We refer to this second group of participants as "\textit{Lay users}". As indicated in section 1, our choice of recruiting two different sets of users is necessary to answer RQ3 \& RQ4 and is situated in line with prior work of \cite{doshi2017towards}. Finally, before data collection we preregistered our design and data analysis plan\footnote{\url{https://bit.ly/apptrust-userstudy}}. With our study design, our aim was to closely replicate the decision-making contexts of the real world police teams by closely simulating the real world conditions. 

\subsection{Designing AI system's Explanations}\label{explanations}
We conducted a preliminary study with three expert users to inform us about the design of explanations (refer footnote 5). Building on insights from this preliminary study, we sought to design our explanations. Also, we added weather and escape route information to the explanation based on an insight which emerged during our preliminary study. Once we decided on the content of the explanations, we looked at established guidelines in the literature on crafting them. We selected the guidelines by Szymanski et al. \cite{szymanski2021visual} because they conducted a state-of-the-art literature survey and a formative study on XAI. The guidelines include (a) quantifying each parameter’s contribution to prediction, (b) what parameters lead to predictions, (c) instances with similar predictions, (d) locating regions about uncertainity, and (e) displaying an overall predictions. To explain the predictive policing system's decision in a textual form, we generated sentences per input parameter using the template described by Hohman et al. \cite{hohman2019telegam}: \textit{The system predicts a higher likelihood of incidents in hotspot A/ B/ C/ D based on (historical crime data OR proximity to dense forest/ highway/ sea OR last arrest of offenders) \textbf{[A]}. The X\% confidence score reflects (strong/ weak) support \textbf{[B]}, and the remaining X\% acknowledges potential unknowns like X \textbf{[C]}. Major contributing factors to this decision include C1 (X\%), C2 (X\%), and C3 (X\%) \textbf{[D]}. Furthermore, a similar case was found in X's police records, where offenders were caught near X \textbf{[E]}. A strong/ no correlation with severe weather (snow or thunderstorms) was found while making this decision \textbf{[D]}. (Tip: Weather prediction for the next three days is X; allocate resources accordingly.}

\begin{figure*}[]
\centering
\includegraphics[width=0.7\textwidth]{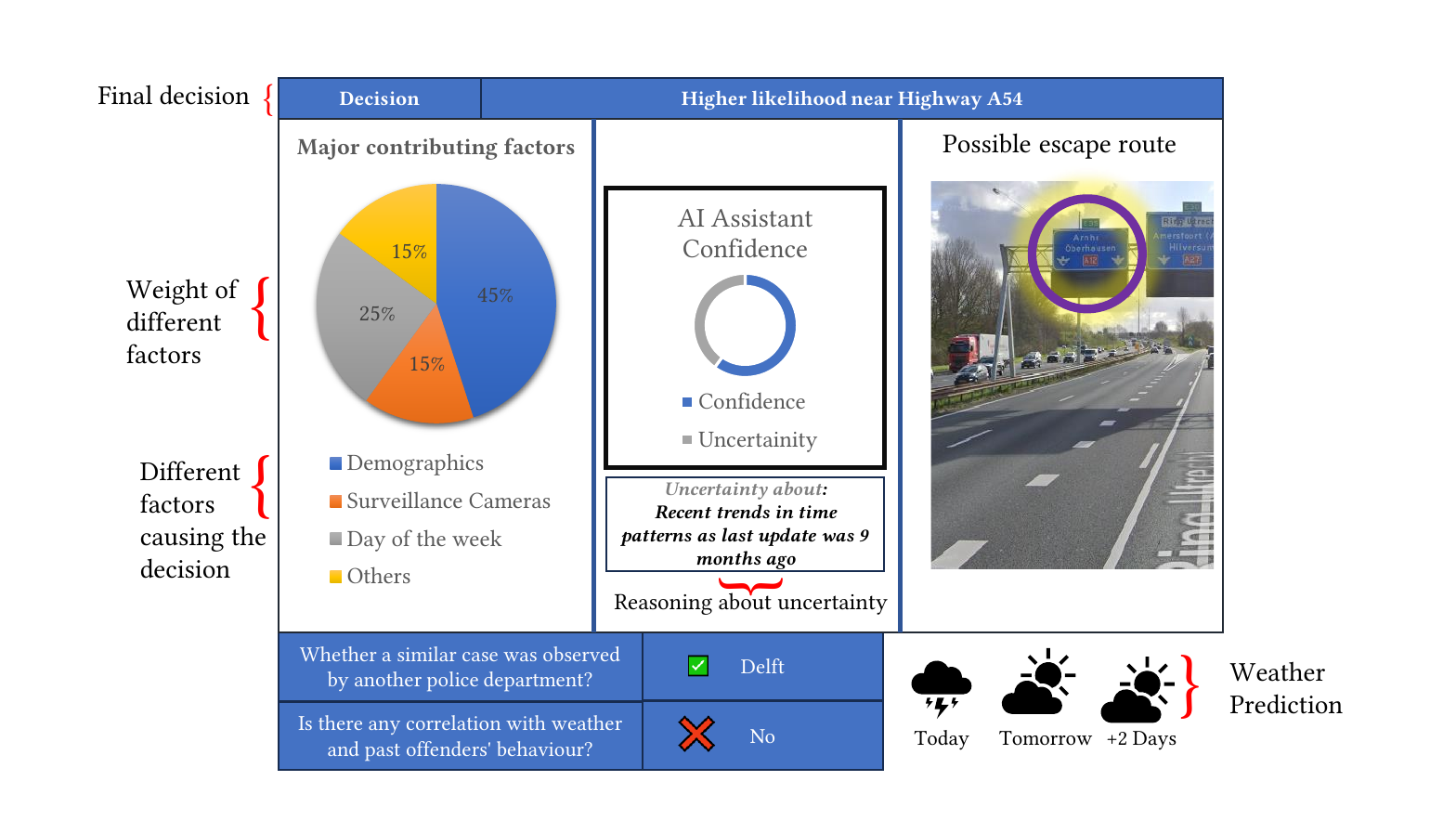} 
\caption{Visual explanations for a selected instance from the user study.}
 \label{explainratings}
\end{figure*}

Here, \textbf{[A]} denotes overall prediction, \textbf{[B]} denotes the confidence in the prediction, and \textbf{[C]} shows regions where the model prediction was uncertain. \textbf{[D]} quantifies each parameter's contributions, and the name of the parameters, and \textbf{[E]} denotes instances that have similar predictions. The contributing factors were among the following: (C1) Historical crime data, (C2) Geographical information, (C3) Time and day of the week, (C4) Weather information, (C4) Demographic statistics, (C5) Resource availability, and (C6) Socioeconomic data. To enable a fair comparison, the visual explanations contained the same information as the textual explanations. Figure \ref{explainratings} shows an example of a visual explanation used in the study. Finally, as prior research on designing hybrid explanations is limited \cite{Mohseni2018ASO}, we based our design only on previous work done by \cite{szymanski2021visual}, who combined visual explanations with text.

\noindent\textbf{Traditional Investigation Notes:}\label{notes}
Based on our preliminary study, police officers often follow traditional methods (diary notes, intel from other units and department instructions) for investigation in conjunction with predictive policing systems. On the one hand these notes serves as the ecological validity (in real life police officers often use their diary notes for investigation) for the task and on the other hand they make sure that there is a 'joint' knowledge for our both groups of participants. An example of a note used in this study is as follows:

\textit{You have (less than a year OR more than three years of experience) in this area shown on the map \textbf{[A]}. According to your diary notes, under the cover of darkness, the past offenders often slip through the labyrinth of narrow alleyways matching with the hotspot A/ B/ C/ D \textbf{[B]}. According to the intelligence department of the Police, the last fugitive vanished into the dense forest after following the alleyways \textbf{[C]}.} Here, \textbf{[A]} denotes the overall experience of the police officer in the selected area, \textbf{[B]} denotes diary notes and probable hotspot selection, and \textbf{[C]} shows the intel from the intelligence department. 
\section{First User Study - "Expert Users"}
\subsection{Participants}
For our expert study, we recruited 12 retired police officers (aged between 65 and 70, 10M:2F) who retired in the last five years from the Dutch police. Our goal was to recruit experts who had prior experience with predictive policing systems or were in-charge of making decisions related to crime prevention. The retired police officers fulfilled these criteria as they were mostly in higher positions in their hierarchy before they retied. Furthermore, given the discussions around the new EU AI Act, police officers in the current force were hesitant to join our user study. Therefore, we decided to recruit retired police officers as they fit our goal of expert users. This study was approved by the Human Research Ethics Review Board of our university and was conducted in the Dutch language.

\subsection{Methodology}\label{study1method}
\subsubsection{Independent variable}\label{explanationtypes}

    \textbf{Explanations} (\textit{categorical, between-subjects}). We assigned each participant to one of four configurations:
    (1) \textbf{No explanation}: participants saw hotspot selection by the AI assistant but not how this decision was made.
        (2) \textbf{Text-based Explanations}: participants saw the hotspot selection by the AI assistant and how this decision was made in textual form.
          (3) \textbf{Visual Explanations}: participants saw the hotspot selection by the AI assistant and how this decision was made in visual form.
         (4) \textbf{Hybrid (Text+Visual) Explanations}: participants saw a combination of text-based and visual explanation.

\subsubsection{Dependent variables.}

     (a) \textbf{Appropriate Trust} (\textit{continuous}). Adapted from \cite{yang2020visual,mehrotraintegrity,zhang2020effect,laiv}. These measures are described in section \ref{apptrustmeasure}.
     (b) \textbf{Subjective Trust} (\textit{continuous}). We used a self-reported global trust meter that captures changes in trust for each round ranged from completely distrust (-100) to completely trust (+100), adapted from \cite{mehrotraintegrity,khasawneh2003model,yang2020visual}.  
    (c) \textbf{Usefulness of Explanations} (\textit{continuous}). The usefulness of explanations was measured on a 7-point Likert scale from Not at all helpful (1) to Very helpful (7), adapted from \cite{yang2020visual}.

\subsubsection{Descriptive and exploratory measurements} We use these variables to describe our sample and for exploratory analyses, but we do not conduct any conclusive hypothesis tests on them.\label{explorevariables}

    (a) \textbf{Age group} (\textit{categorical}). Participants will select their age group from multiple choices.
    (b) \textbf{Level of education} (\textit{categorical}). Participants will select the highest level of education they have completed.
    (c) \textbf{AI literacy} (\textit{continuous}). Average score of the four items defined by \cite{schoeffer2022there}.
    (d) \textbf{Propensity to Trust} (\textit{continuous}). Propensity to Trust scale by Merritt et al. \cite{merritt2011affective} adapted to understand apriori trust in the predictive policing systems.
    (e) \textbf{Personal experiences as a police officer and use of AI} (1)) Have you ever worked with predictive policing systems in the past? and (2)) Do you have prior experience with the use of AI in predictive policing systems?
    (f) \textbf{Task stakes perception} (\textit{continuous}). In this study we have considered scenario such as pick-pocketing as non-violent crime and sexual-offense as a violent crime based on the Dutch WODC Magazine Recidivism\footnote{\url{https://magazines.wodc.nl/wodcmagazine/2019/03/high-impact-crime-hic}}. Since the stakes involved in a decision are subjective \cite{shivani}, we will capture task stakes perceptions using \cite{lyons2022}.
    (g) \textbf{AI Confidence Score} (\textit{categorical, within-subjects}) AI accuracy was communicated to participants as a part of the explanations. (1) High (\textit{Confidence Score $>$ 75\%}) and (2) Low (\textit{Confidence Score $<$ 75\%}). 
       (h) \textbf{Geographical Experience} (\textit{categorical, within-subjects}): Prior experience with policing about the shown geographic area on the map was communicated in the dairy notes. (1) Amount of professional experience : Limited (\textit{$>$ 3 years experience}) and Amount of professional experience : High (\textit{$<$ 3 years experience}). 

\subsubsection{Measurement of appropriate trust}\label{apptrustmeasure}
In this study, we used two measurements of appropriate trust \cite{yang2020visual,mehrotraintegrity} and calibrated trust \cite{laiv,zhang2020effect} each from prior research. We used distinctive measures for appropriate and calibrated trust based on the definitions provided in the literature \cite{wischnewski2023measuring,okamura2020adaptive,de2020towards}. For example, Mehrotra et al. show that different definitions and measures of appropriate and calibrated trust exist in the literature \cite{mehrotra2023systematic}. 
We argue that it is necessary to study multiple measures to understand when trust can be classified as appropriate or calibrated, as different measures may result in slightly different outcomes. Also, it is important to differentiate between trust, trustworthiness and reliance as illustrated by Tolmeijer et al \cite{suzanne2022capable}. In this paper, we therefore define trust as the belief that “an agent will help achieve an individual’s goal in a situation characterized by uncertainty and vulnerability” \cite{leesee}, trustworthiness as an antecedent of trust \cite{mayer1995integrative} and reliance  as “a discrete process of engaging or disengaging” \cite{leesee} with the AI system. Our measures are:

\noindent\textbf{Measure 1 (App1):} Appropriate trust is to [not] follow an [in]correct recommendation \cite{yang2020visual}\footnote{As opposed to Yang et al \cite{yang2020visual}, we will treat this measure to appropriate reliance as this measure only focuses on engaging or disengaging with the AI system.}.
    
\noindent\textbf{Measure 2 (App2):} Appropriate trust occurs when (a) the human estimates that the AI agent is better at the task than the human, (b) also the actual Trustworthiness\footnote{Here, actual trustworthiness is an inherent characteristic of a system that subsumes true capabilities of the system in question \cite{Schlicker}.} (TW) of the AI agent is equal to or higher than the human’s TW and (c) the human selects the AI agent for the task and \textit{vice-versa} \cite{mehrotra2023building}.
    
\noindent\textbf{Measure 3 (Calib1):} Switch percentage, the percentage of trials in which the participant switched from their initial prediction to use the AI’s prediction as their final prediction \cite{zhang2020effect}. 
   
\noindent\textbf{Measure 4 (Calib2):} Agreement percentage, the percentage of trials in which the participant’s final prediction agreed with the AI’s prediction \cite{laiv}.
\begin{table}[h]
\centering
\caption{Categorization of measures of appropriate trust}
\label{tab:example}
\begin{tabular}{|c|c|c|c|c|}
\hline
\textbf{Round} & \textbf{Human TW} & \textbf{AI TW} & \textbf{AI} & \textbf{Human} \\
\hline
1 & High & High & Correct & Correct \\ 
2 & Low & High & Correct & Incorrect \\ 
3 & Low & Low & Correct & Incorrect \\ 
4 & High & High & Incorrect & Correct \\ 
5 & High & Low & Correct & Correct \\ 
6 & Low & High & Incorrect & Incorrect \\ 
7 & Low & Low & Incorrect & Correct \\ 
8 & High & Low & Incorrect & Correct \\
\hline
\end{tabular}
\end{table}

In \textbf{App2}, (a) is cognitive trust from the human or the perceived trustworthiness of the AI system, (b) is
classification of trustworthiness based on Table \ref{tab:example} and (c) is human selection that could be based on observable behaviour, rationality or simply delegation of the responsibility. For the classification of actual trustworthiness, we have divided the human expertise based on the traditional investigation notes (\textit{refer Section 3.2}) as High or Low and AI expertise based on the available location data as High or Low. Furthermore, we have categorized the cases where AI's suggestion and human's diary notes correspond to the correct selection of the hotspot for predictive policing. The hotspot’s correctness can never be proven in real life as it would require complete information about all the areas. Hence, to simplify, in our work the correctness of a hotspot was simply based on the permutations of trustworthiness and combinations of AI \& Human correctness, refer to Table \ref{tab:example}. As to \textbf{Calib1} and \textbf{Calib2}, the main difference between these two measures of calibrated trust was the agreement percentage. \textbf{Calib2} would count the trials in which the participants and the AI's predictions agreed and counted as the final decision. In contrast, the switch percentage \textbf{Calib1} would only consider cases where they disagreed and had to switch intentionally.  

%\subsubsection{Procedure}
%\begin{figure*}[]
%  \centering
%  \includegraphics[width=\linewidth]{Expflow.pdf}
%  \caption{The experimental design of the user study. Each participant was assigned to a specific explanation type (Baseline - no explanation, Text, Visual and Hybrid) and they finished 8 rounds.}
%  \label{expflow}
%\end{figure*}
Our experiment aimed for human-in-the-loop collaboration, where participants made decisions assisted by an AI assistant. Participants, after providing informed consent and answering descriptive and exploratory questions from section 4.2.3, were introduced to the AI assistant. The AI assistant was hypothetically trained on the last ten years' crime data in the Netherlands. Finally, they were assigned to one of the four explanation types as per section \ref{explanationtypes}.

\noindent\textbf{Step 1}: \textit{Trial Study} - Participants were tasked with choosing a hotspot for resource allocation, specifically for police patrol. They read investigation notes for guidance and made their selection. The AI assistant then chose a hotspot based on hypothetical training data from \url{https://data.politie.nl/}, offering reasoning using one of four explanation configurations. Participants were then prompted to make a final selection, either affirming their or the AI assistant's choice or selecting a different hotspot and providing a reason.
\begin{figure*}[h]
  \centering
\includegraphics[width=\linewidth,height=\textheight,keepaspectratio]{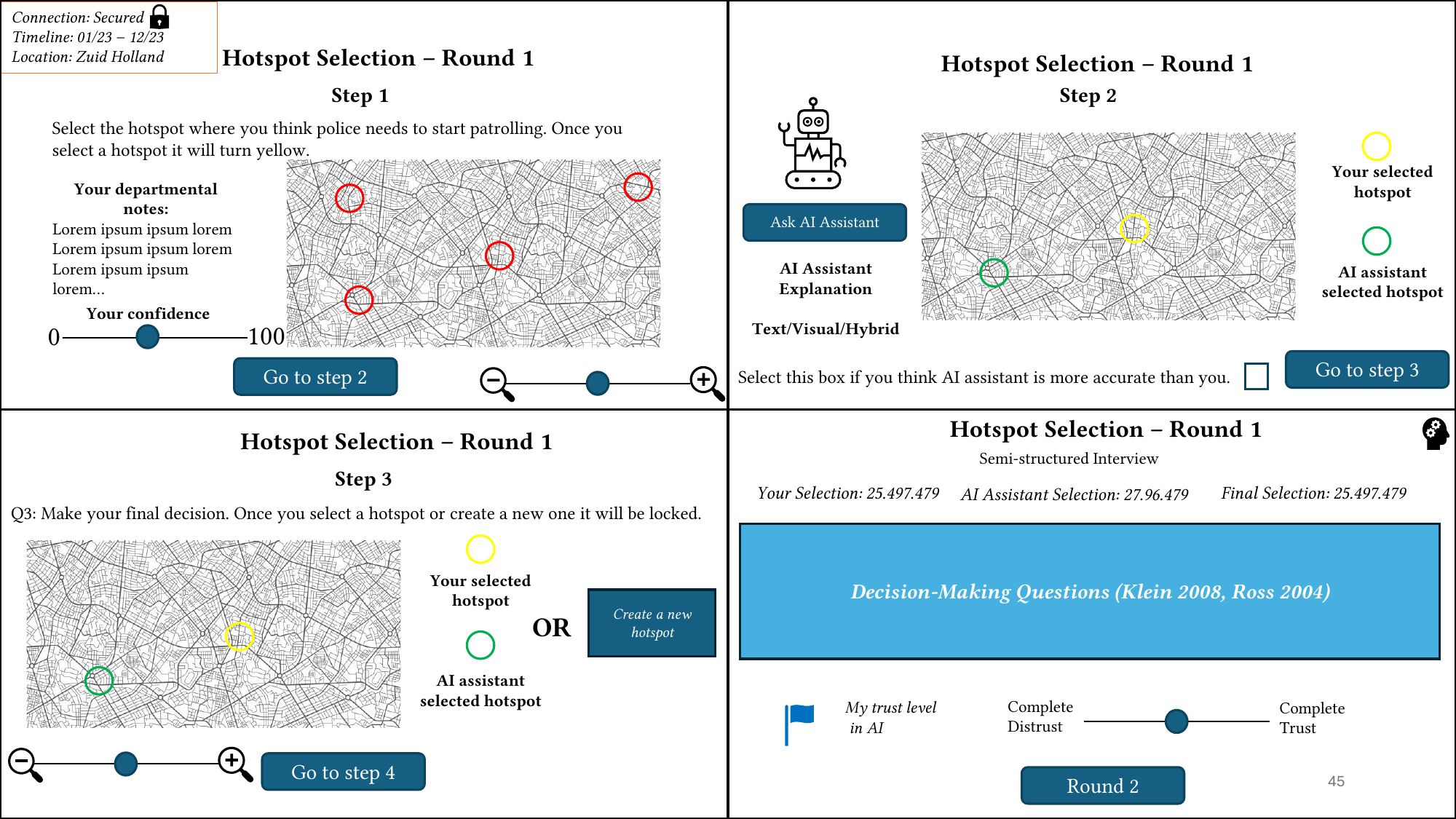}
  \caption{An illustration of the four steps performed by participants of the user-study. In step 1, participants rate their confidence in accurately identifying the hotspot. In step 2, the AI assistant selects a hotspot with its reasoning in form of explanations. In step 3, the participants makes their final decision (Q3). Finally, in step 4, participants rate their subjective trust, usefulness of the explanations and open questions related to decision-making.}
  \label{flowofstudy}
\end{figure*}

\noindent\textbf{Step 2}: \textit{Main Study} - After completing the trial round, the participants received details about the main study, which consists of eight rounds. The participants had a limited 3-minute window for hotspot selection informed by our preliminary study with expert users. In addition, at this step, participants were asked to tick a checkbox if they believed that the AI assistant was better at the task than they themselves. They were instructed that the (\textit{hypothetical}) intelligence unit chief would review their hotspot selections at the end of the experiment. Correct selections earned +10 points, while incorrect ones incurred a -10 point deduction. Due to time constraints, immediate result verification was not possible, prompting participants to proceed to the next round swiftly. Additionally, a bonus was promised for those achieving a top 3 score. Figure \ref{flowofstudy} describes all the steps performed by the participants.

\noindent\textbf{Step 3}: \textit{End of the Study} - Participants completed the post-experiment questionnaire after the main study. They were asked to rate task stakes perception and how familiar they were with the geographical areas shown in the study. Furthermore, we also asked them two open-ended questions: (a) Do you think the AI assistant offered an appropriate explanation of the decision-making process? Why? If not, what explanation do you think it should have offered? and (b) How was your overall experience with this user study? Once they answered these questions, they were shown their final score. A pilot study with five HCI researchers revealed no significant usability flaws, see footnote 5 for details.

\begin{figure}[hbt!]
  \centering
\includegraphics[width=\linewidth,height=\textheight,keepaspectratio]{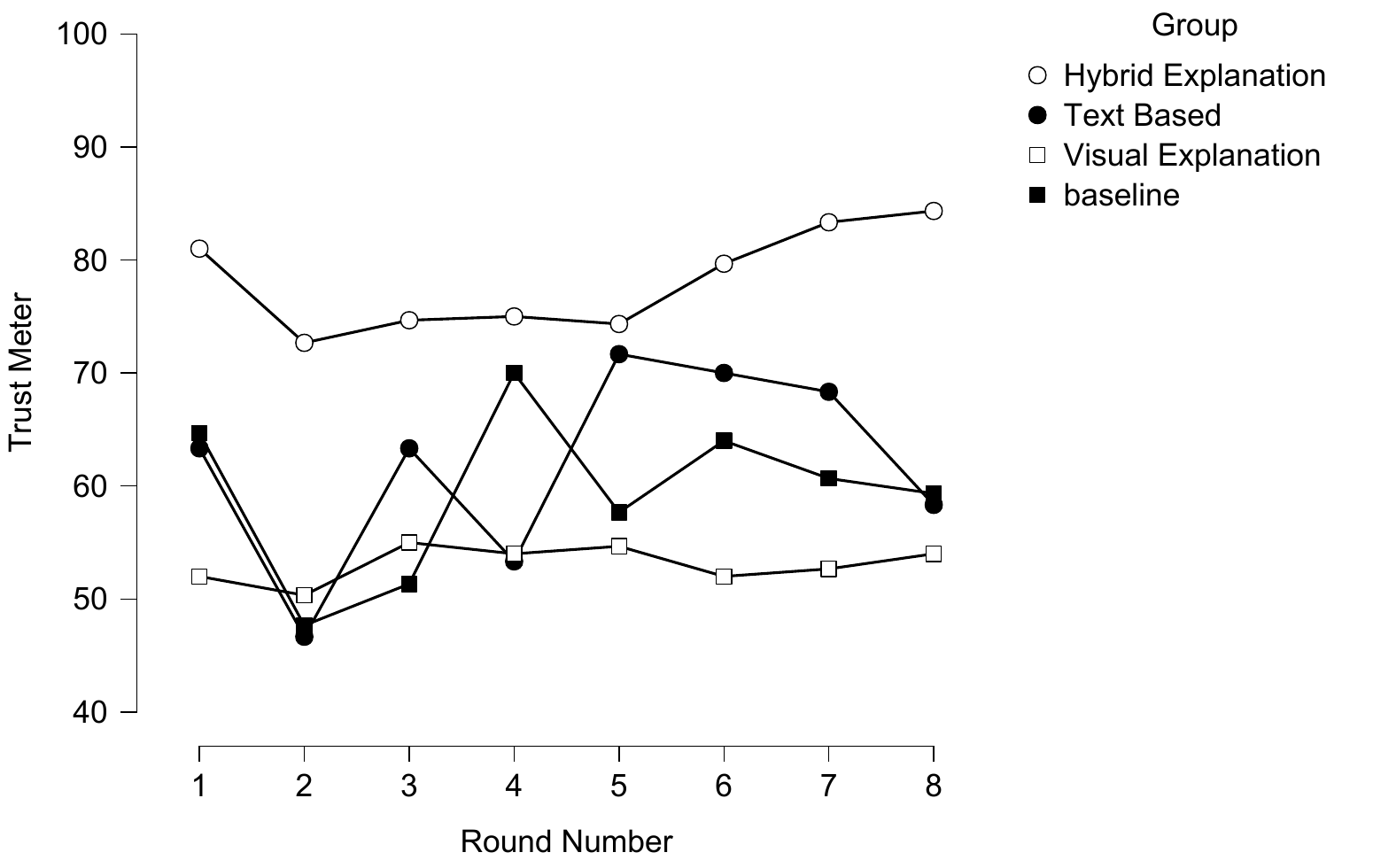}
  \caption{An illustration of mean responses for changes in Global Trust Meter over 8 rounds for study 1. }
  \label{subtrutNLratings}
\end{figure}

\subsection{Results}
\subsubsection{Descriptive Statistics}
Of the 12 participants in our user study, 10 had at least a Bachelor's degree. All of them claimed to have know about predictive policing systems, whereas only two of them had heard of or had experience using AI in predictive policing. Our participants' average score of AI literacy was 6.8 (SD = $\pm$1.1) on a scale of 0 to 10, and their propensity to trust predictive policing system was 7.2 (SD = $\pm$1.25) on a scale of 0 to 10. This apriori trust in the system highlights that interface design issues are important to consider as participants in general considered the system legitimate and trustworthy. %The average duration of the study was 32 minutes (SD = $\pm$3.25), and each participant spent an average of 2 minutes 35 seconds (SD = $\pm$0.60) per round. 

\noindent Our analysis revealed no major differences in the frequency count of our measures of appropriate trust between the explanation types. However, subjective trust scores were comparatively higher for hybrid explanations (Mean = 60.98, SD = 5.62) when compared to all other explanation types (Mean = 41.23, SD = 10.83), see Figure \ref{subtrutNLratings}.

\subsubsection{Qualitative Analysis}\label{qualstudy1}
We first translated the transcripts in English with the help of two native bilingual speakers and performed qualitative analysis using a reflexive thematic analysis \cite{braun2006using}. We inductively generated individual codes from our participants' responses to the open-ended questions and then clustered them into code groups. We identified two main areas: one related to explanation presentation \& clarity, and the other related to perceptions of the AI system.

\noindent\textbf{Explanations Presentation and Clarity:} Overall, participants from the hybrid and textual explanations group found the explanations to be clear and structured. P7 (textual explanation) wrote, ``\textit{the use of public language rather than technical jargon helped decide to go with the AI assistant}". On the other hand, there were mixed reviews from participants for the visual explanations and no explanations categories. P11 (no explanations) expressed the desire for more underpinning or context with the explanation. P5 (visual explanation) found visual explanations to be overwhelming. Also, 50\% of participants wrote they followed their reasoning first and then looked at AI assistant's recommendation.

\noindent\textbf{Perceptions of the underlying AI system:} There were mixed reviews regarding the help provided by the AI assistant. From the far opposing end, there were concerns about the use of AI in predictive policing where P8 (baseline) wrote, ``\textit{the use of AI in predictive policing is fundamentally wrong because you cannot train a system to do policing}". Interestingly, we also found some quotes related to AI capabilities that supported P8's thinking such as ``\textit{even explanations will not help in trusting fundamentally biased system}". Also, P8's apriori trust in system was rated at 2 on a scale of 0 to 10, which also supports P8's views. Of the two other participants, who had a rating of 5 or less wrote, ``\textit{AI does not possess human intuition and experience. Hence it cannot help in the way that my notes from my teams can"} - P12 (textual)  and ``\textit{AI rarely captures the considerations of the perpetrator, which is important in understanding any crime as found discussed among officers"} - P4 (Hybrid). Some participants appreciated the AI assistant's decision, when it aligned with their understanding (P1, P7, P10, P15) or provided additional information that was new to them. For example, P1 wrote, ``\textit{The information about the existing military unit was useful because it requires a cooperative operation then.}"
\section{Second User Study - "Lay Users"}
We conducted another user study to understand user expertise's role and assess explanations' role in fostering appropriate trust at a scale. We computed the required sample size using G*Power \cite{faul2009statistical} for an ANOVA with main effects and interactions, specifying the default effect size of 0.25, a significance threshold of $\alpha$ = 0.05, a desired power of 0.8, four groups, and the respective degrees of freedom. The result indicated that we require approximately 179 participants. 

We recruited 209 participants from \textit{Prolifc} with an approval rate greater than 95\%. Each participant was at least 18 years old, highly proficient in English, and could participate in our study only once. Participants were rewarded based on a \$10 hourly rate, and the median completion time was 28 minutes and 11 seconds. Participants were excluded from data analysis if they did not pass at least one of the attention checks. This led to 180 participants (age between 18 and 65+, 94M:86F), \textit{i.e.,} 45 participants per explanation type. The study was conducted on \textit{Qualtrics} in English and was approved by our university's review Board.

\subsection{Methodology}
Each participant had to follow the same methodology as the first user study except that they also answered the following question - Have you worked for the police in the past, or are you currently working? Concretely, this means we wanted to filter the participants who have worked for the police in the past or the present as they might have expert knowledge about predictive policing and would not classify as lay users.

\subsection{Results}
\subsubsection{Descriptive Statistics}
Of the 180 participants in our user study, 29.44\% were between 18 and 24 years old, 44.44\% between 25 and 34 years old, 17.77\% between 35 and 44 years old, and 8.33\% were between 45-65+. 77\% of the participants had at least a Bachelor’s degree. None of them claimed neither to have ever worked for the police nor were they aware of the predictive policing system. The average score of AI literacy among participants was 4.67 (SD = $\pm$1.25), and their propensity to trust AI systems was 4.31 (SD = $\pm$0.91). The average duration of the study was 23 minutes (SD = $\pm$4.25), and each participant spent an average of 2 minutes 5 seconds (SD = $\pm$1.03) per round to make the final selection.

\subsubsection{Inferential Statistics}
Before conducting any statistical analyses, we mapped all (seven-point) Likert scale answers onto an ordinal scale ranging from  - 3 (i.e., strongly disagree) to 3 (i.e., strongly agree). The result of Shapiro-Wilk shows that our data followed the normal distribution. Therefore, we conducted an ANOVA with explanations as between-subjects factors and different measures of appropriate trust as the dependent variables. Next to the \textit{F} statistic and \textit{p}-value, we also report the partial eta squared $\eta_p^{2}$ effect size. We found no main effect of different explanation types on any measure of appropriate trust (\textit{p} $>$ 0.05, $\eta_p^{2}$ $<$ 0.01, cohen's \textit{d} $=$ 0.65).

We conducted another ANOVA with the same between-subjects factors but with subjective trust ratings and usefulness of explanations as the dependent variable. We found a significant difference between different explanation types on the perceived usefulness of the explanations (F (3,1436) = 4.35, \textit{p} $<$ 0.005, $\eta_p^{2}$ = 0.2, \textit{d} $=$ 0.71). The post hoc analysis revealed that hybrid (\textit{p} $<$ 0.013) and visual explanations (\textit{p} $<$ 0.001) were significantly better than no explanations for the usefulness of explanations ratings. However, we did not find any evidence indicating the effect of different explanation types on subjective trust responses (\textit{p} = 0.479, $\eta_p^{2}$ $<$ 0.01).

In addition to the analyses described above, we conducted multiple linear regression to analyze the association of independent and dependent variables and exploratory analyses to explore any trends in the data. Our results show that \textbf{App2} ($\beta$ = 4.31, \textit{p} $<$ 0.001), \textbf{Calib2} ($\beta$ = 8.59, \textit{p} $<$ 0.001) and AI assistant's correctness ($\beta$ = 11.01, \textit{p} $<$ 0.001)  predicted the measure \textbf{App1} (\textit{$R^2$} = 0.40, AIC = 1502, BIC = 1560), with AI assistant's correctness being the strongest predictor. Similarly, we found perceived usefulness of explanations ($\beta$ = 2.16, \textit{p} $<$ 0.001) and AI assistant trustworthiness ($\beta$ = 12.58, \textit{p} $<$ 0.001) predictors of \textbf{App2} (\textit{$R^2$} = 0.393, AIC = 1311, BIC = 1314) other than \textbf{App1}. We also found \textbf{Calib2} ($\beta$ = 9.34, \textit{p} $<$ 0.001) and perceived usefulness of explanations ($\beta$ = 19.45, \textit{p} $<$ 0.001) predicted the subjective trust scores (\textit{$R^2$} = 0.267). Finally, we did not find evidence of any exploratory variable affecting measures of appropriate trust. 

\subsubsection{Qualitative Analysis}
We followed a similar approach as in section \ref{qualstudy1} to perform our qualitative analysis. We identified two main topics of interest:  %This section explains each of those areas in detail. Appendix B shows all selected quotes.

\noindent\textbf{Evaluation of AI's reasoning: } Participants, in general, had a positive attitude towards the AI assistant across all explanation types due to (a) lack of expertise for the task, P24 (no explanation): \textit{``I think this system know what it is going, I just need to use it accordingly as this task is very new to me"}, (b) in-depth reasoning of the decision, P96 (textual explanation): \textit{{``I believe various factors considered by the AI, such as historical crime data, weather, demographics, and spatial relationships are useful to decide."}}, and (c) breaking the tunnel vision, P77 (visual explanation): \textit{{``I find visual information appealing and photos, maps, past crime patterns are right to the point, especially the link with weather is something I could never think off."}} Some participants expressed reassurance from AI's logical reasoning (P9, P23, P55, P149, P180) and expressed higher trust when their hotspot selection was similar to AI (P112, P106, P155, P47, P33).

\noindent\textbf{Doubts about AI's effectiveness: }Several participants (23\%) expressed scepticism about the AI assistant's effectiveness irrespective of explanation types. They put forward the desire for consideration of (a) real-time factors (P25, P40, P164), (b) more transparency (P54, P109, P1172), (c) resolution of discrepancies between AI and personal judgement (P37, P111, P140), and (d) providing validation approaches for AI decision-making (P50, P74, P101). Furthermore, 12 participants reported that if the explanation was hard to understand and follow, they just followed the AI assistant's answer because it is too much work to determine whether the AI is right or wrong. E.g., P78: "\textit{region around Assen is under control of military so how as a police officer can make any judgement, go with AI!"}

\section{Discussion}
\subsection{Effect of explanations on appropriate and subjective trust (\textbf{RQ1 and RQ2})}
Our findings show that explanation types, including `no explanations' had no impact on any measures of appropriate trust in either user study (RQ1). This lack of effect can be understood through the lens of the illusion of explanatory depth \cite{chromik}, a cognitive bias where individuals overestimate their understanding of complex systems. In the context of AI explanations, users could have believed that they comprehend the AI's decision-making process more thoroughly than they actually do, leading to a false sense of understanding regardless of the explanation provided.

To further interpret our results, let's revisit the definitions of our measures, \textbf{App1} and \textbf{App2}. For \textbf{App1} to occur, participants must [not] follow [in]correct recommendations \textit{i.e.,} their appropriate reliance on the system, and for \textbf{App2}, understanding both the trustor and trustee's trustworthiness is crucial. Our analyses indicate that, on average, participants correctly selected the hotspot four times in study 1 and three times out of eight rounds in study 2, suggesting a 50\% error rate. This high error rate could be attributed to explainability pitfalls, where explanations fail to effectively convey the AI system's reasoning or limitations \cite{de2022perils}. 

Moreover, in user study 1, participants utilized the AI assistant to confirm their intuitive professional judgment rather than comparing trustworthiness, leading to a lack of substantial variations in explanatory formats. Therefore, regardless of expertise, participants failed to perceive meaningful distinctions in trustworthiness, leading us to conclude that there was no effect on appropriate trust, regardless of the type of explanation provided.

For RQ2, our findings indicated that hybrid explanations were rated better on subjective trust than all other explanation types in study 1.  However, this trend was not apparent in study 2. This result suggests potential variations in how different user groups perceive and respond to explanation types echoing the work by Szymanski et al. \cite{szymanski2021visual}. These differences can be understood through the lens of the ironies of automation, which posit that as systems become more automated, the role of human operators becomes more critical yet potentially more challenging \cite{endsley2023ironies}. %Several other factors could contribute to this divergence. It is conceivable that the prior professional experience of retired police officers influenced their preference for hybrid explanations, given their familiarity with complex decision-making processes. In contrast, lay users in study 2 might have different expectations or preferences. %Additionally, the significant positive correlation between (a) apriori trust in the system, (b) AI literacy and (c) subjective trust ratings which was only visible for retired police officers could have contributed to the observed differences.

\subsubsection{Increase in trust doesn't mean trust is appropriate}The relationship between increased trust and its appropriateness in AI systems presents a complex challenge that warrants careful examination. Our findings reveal that while hybrid explanations significantly enhanced subjective trust in study 1, this increase did not correlate with improved appropriateness. This disconnect challenges the prevalent assumption that higher trust inherently leads to better outcomes \cite{bansal2021does}, particularly in sensitive domains like predictive policing where the effectiveness of decision-making is paramount.

Appropriate trust represents a multifaceted construct influenced by context, agent characteristics, and underlying cognitive processes \cite{chen2014human}. Our research suggests that while explanations may boost user trust in AI systems, this enhanced trust does not necessarily translate to improved decision-making capabilities. This finding presents two potential paths forward: exploring alternative trust-building approaches or addressing potential deficiencies in explanation quality \cite{bansal2021does}.

Current research presents divergent perspectives on addressing this challenge. Some researchers advocate for diversifying trust-building methods through transparency, user engagement, and iterative feedback \cite{carolina,ulfert2023shaping}, while others emphasize the need to enhance explanation quality and clarity \cite{mehrabi2021survey}. Based on our findings, we recommend a balanced approach that evaluates trust through multiple lenses: appropriateness (goal alignment), system purpose (usability), and user requirements (usefulness). Additionally, we suggest exploring Miller's Evaluative AI framework as a potential alternative to traditional XAI approaches, offering hypothesis-driven decision support \cite{miller2023explainable}. %The author reports that evaluative AI approach helps in trust calibration because (a) it aligns with the human cognitive decision-making process, and (b) there is no recommendation to follow which pushes users not to trust blindly.

\subsection{Usefulness of explanations, role of user expertise and exploratory measures}
Our analysis reveals several key findings regarding the relationship between explanations, trust, and user expertise in AI systems. Perceived usefulness of explanations significantly influenced subjective trust scores (RQ3), though this did not translate to appropriate trust formation. The correlation between explanation usefulness and subjective trust suggests that users' trust assessments incorporate both prediction accuracy and the perceived value of explanations.

While user expertise did not moderate the role of explanations in building appropriate trust (RQ4), expert users showed significantly higher subjective trust with hybrid explanations. This aligns with existing research emphasizing the importance of user expertise in AI system trust \cite{szymanski2021visual,ribera2019can,Mohseni2018ASO} and supports previous arguments for expertise-tailored explanations \cite{arrieta2020explainable,burnett2020explaining,sperrle2020should,gunning}. A marked distinction emerged in decision-making patterns between user groups. Lay users predominantly followed AI recommendations directly (42\%), while expert users demonstrated a more analytical approach (75\%) through their open-text responses. This difference aligns with Wang et al.'s findings \cite{wang2019designing} regarding inexperienced users' susceptibility to reinforcement effects, particularly evident when lay users encountered unfamiliar information.

Our exploratory variables analysis revealed positive correlations between AI trust propensity, AI literacy, subjective trust scores, and crime classification in study 1. However, these correlations did not persist in study 2, suggesting the influence of contextual factors. This variance might be attributed to differences in participant characteristics (study 1 showed higher average AI literacy scores of 6.80 compared to 4.67 in study 2) and system-specific factors in predictive policing. The identified predictors of appropriate trust, particularly AI assistant correctness and trustworthiness, demonstrated consistency across both expert and lay user contexts.

\subsubsection{Policy measures for public sector AI systems}

Public sector AI systems, particularly in predictive policing, face critical temporal and spatial challenges that require careful consideration before development or deployment. Our research identifies three key challenges: confirmation bias in AI interpretation across both expert and non-expert users, alignment with pre-existing biased judgments which risks amplifying systemic biases \cite{selten}, and the concerning disconnect between increased subjective trust and actual decision quality when explanations are provided.

These findings carry significant policy implications for responsible AI implementation. Most notably, the observation that increased subjective trust through explanations does not correlate with improved decision-making necessitates a strategic shift in AI development approaches. We propose three essential policy measures that extend to broader XAI systems: implementing performance metrics focused on decision quality rather than subjective trust; incorporating user expertise levels into system design, given the observed variations between user groups; and preserving human discretion in decision-making, as evidenced by expert users' successful integration of AI advice with professional knowledge. This last point is particularly relevant when compared to systems like Germany's, where deviation from AI recommendations is constrained \cite{meijer2021algorithmization}.

\subsection{Limitations and Future Work}
Our study faced several notable limitations that should inform the interpretation of results and future research directions. The primary constraint was Study 1's small sample size, which, while consistent with sampling methods used in comparable predictive policing research \cite{gerber2017justifying}, potentially limits the generalizability of our findings to broader populations. 
The study's experimental design, utilizing hypothetical scenarios focused on individual hotspot selection with AI assistance, represents a simplified version of actual police decision-making processes, which typically involve team-based approaches. This aligns with Ferguson's observations \cite{ferguson2018legal} regarding "big data policing," where concerns often center more on underlying data quality than system trust. The use of experimental conditions and fictitious vignettes, while methodologically necessary, couldn't fully capture the complexity of authentic police decision-making environments.
To address these limitations, future research should employ larger sample sizes and explore alternative XAI approaches to better understand appropriate trust development. More robust methodological approaches could include virtual reality simulations of police decision-making \cite{mcdaniel2021predictive}, deliberative polling techniques, or real-world interventions. These enhanced approaches would provide more comprehensive insights into the complexities of trust formation in predictive policing systems.
\section{Conclusion}
In this paper, we looked at the effect of different types of explanations (text, visual, and hybrid) and user expertise (retired police officers and lay users) on fostering appropriate trust in an AI-based predictive policing system. Our results show that a hybrid form of explanations raised the subjective trust of expert users compared to lay users in the AI system. However, none of the explanation types helped participants in forming appropriate trust in the system. We argue that this result of an increase in trust is worrisome, as it does not lead to better decisions. Based on these results, we highlight challenges in building appropriate trust in human-AI interaction and propose important policy recommendations centered around fostering appropriate trust in AI-based predictive policing systems. We hope this paper will serve as a “call to action” for the UMAP community to shift focus from the use of explanations for just promoting trust in AI systems to fostering appropriate trust instead.
\section{Acknowledgements}
This research was (partly) funded by the Hybrid Intelligence Center, a 10-year programme funded the Dutch
Ministry of Education, Culture and Science through the Netherlands Organisation for Scientiic Research, grant
number 024.004.022, by TAILOR Connectivity Fund Grant through EU H2020 program under agreement number 952215 and by EU H2020 ICT48 project ``Humane AI Net" under contract 952026.
\bibliographystyle{ACM-Reference-Format}
\bibliography{aaai24}


\begin{thebibliography}{90}

%%% the Web address.

\ifx \showCODEN    \undefined \def \showCODEN     #1{\unskip}     \fi
\ifx \showDOI      \undefined \def \showDOI       #1{#1}\fi
\ifx \showISBNx    \undefined \def \showISBNx     #1{\unskip}     \fi
\ifx \showISBNxiii \undefined \def \showISBNxiii  #1{\unskip}     \fi
\ifx \showISSN     \undefined \def \showISSN      #1{\unskip}     \fi
\ifx \showLCCN     \undefined \def \showLCCN      #1{\unskip}     \fi
\ifx \shownote     \undefined \def \shownote      #1{#1}          \fi
\ifx \showarticletitle \undefined \def \showarticletitle #1{#1}   \fi
\ifx \showURL      \undefined \def \showURL       {\relax}        \fi
% invisible to TeX
\providecommand\bibfield[2]{#2}
\providecommand\bibinfo[2]{#2}
\providecommand\natexlab[1]{#1}
\providecommand\showeprint[2][]{arXiv:#2}

\bibitem[Adadi and Berrada(2018)]%
        {adadi2018peeking}
\bibfield{author}{\bibinfo{person}{Amina Adadi} {and} \bibinfo{person}{Mohammed Berrada}.} \bibinfo{year}{2018}\natexlab{}.
\newblock \showarticletitle{Peeking inside the black-box: a survey on explainable artificial intelligence (XAI)}.
\newblock \bibinfo{journal}{\emph{IEEE access}}  \bibinfo{volume}{6} (\bibinfo{year}{2018}), \bibinfo{pages}{52138--52160}.
\newblock


\bibitem[Akpinar et~al\mbox{.}(2021)]%
        {akpinar2021effect}
\bibfield{author}{\bibinfo{person}{Nil-Jana Akpinar}, \bibinfo{person}{Maria De-Arteaga}, {and} \bibinfo{person}{Alexandra Chouldechova}.} \bibinfo{year}{2021}\natexlab{}.
\newblock \showarticletitle{The effect of differential victim crime reporting on predictive policing systems}. In \bibinfo{booktitle}{\emph{Proceedings of the 2021 ACM conference on fairness, accountability, and transparency}}. \bibinfo{pages}{838--849}.
\newblock


\bibitem[Alikhademi et~al\mbox{.}(2022)]%
        {alikhademi2022review}
\bibfield{author}{\bibinfo{person}{Kiana Alikhademi}, \bibinfo{person}{Emma Drobina}, \bibinfo{person}{Diandra Prioleau}, \bibinfo{person}{Brianna Richardson}, \bibinfo{person}{Duncan Purves}, {and} \bibinfo{person}{Juan~E Gilbert}.} \bibinfo{year}{2022}\natexlab{}.
\newblock \showarticletitle{A review of predictive policing from the perspective of fairness}.
\newblock \bibinfo{journal}{\emph{Artificial Intelligence and Law}} (\bibinfo{year}{2022}), \bibinfo{pages}{1--17}.
\newblock


\bibitem[Arrieta et~al\mbox{.}(2020)]%
        {arrieta2020explainable}
\bibfield{author}{\bibinfo{person}{Alejandro~Barredo Arrieta}, \bibinfo{person}{Natalia D{\'\i}az-Rodr{\'\i}guez}, \bibinfo{person}{Javier Del~Ser}, \bibinfo{person}{Adrien Bennetot}, \bibinfo{person}{Siham Tabik}, \bibinfo{person}{Alberto Barbado}, \bibinfo{person}{Salvador Garc{\'\i}a}, \bibinfo{person}{Sergio Gil-L{\'o}pez}, \bibinfo{person}{Daniel Molina}, \bibinfo{person}{Richard Benjamins}, {et~al\mbox{.}}} \bibinfo{year}{2020}\natexlab{}.
\newblock \showarticletitle{Explainable Artificial Intelligence (XAI): Concepts, taxonomies, opportunities and challenges toward responsible AI}.
\newblock \bibinfo{journal}{\emph{Information fusion}}  \bibinfo{volume}{58} (\bibinfo{year}{2020}), \bibinfo{pages}{82--115}.
\newblock


\bibitem[Bansal et~al\mbox{.}(2021)]%
        {bansal2021does}
\bibfield{author}{\bibinfo{person}{Gagan Bansal}, \bibinfo{person}{Tongshuang Wu}, \bibinfo{person}{Joyce Zhou}, \bibinfo{person}{Raymond Fok}, \bibinfo{person}{Besmira Nushi}, \bibinfo{person}{Ece Kamar}, \bibinfo{person}{Marco~Tulio Ribeiro}, {and} \bibinfo{person}{Daniel Weld}.} \bibinfo{year}{2021}\natexlab{}.
\newblock \showarticletitle{Does the whole exceed its parts? the effect of ai explanations on complementary team performance}. In \bibinfo{booktitle}{\emph{Proceedings of the 2021 CHI Conference on Human Factors in Computing Systems}}. \bibinfo{pages}{1--16}.
\newblock


\bibitem[Barbosa et~al\mbox{.}(2022)]%
        {barbosa2022investigating}
\bibfield{author}{\bibinfo{person}{Gabriel Diniz~Junqueira Barbosa}, \bibinfo{person}{Dalai dos Santos~Ribeiro}, \bibinfo{person}{Marisa do Carmo~Silva}, \bibinfo{person}{H{\'e}lio Lopes}, {and} \bibinfo{person}{Simone Diniz~Junqueira Barbosa}.} \bibinfo{year}{2022}\natexlab{}.
\newblock \showarticletitle{Investigating the relationships between class probabilities and users’ appropriate trust in computer vision classifications of ambiguous images}.
\newblock \bibinfo{journal}{\emph{Journal of Computer Languages}}  \bibinfo{volume}{72} (\bibinfo{year}{2022}), \bibinfo{pages}{101149}.
\newblock


\bibitem[Bayer et~al\mbox{.}(2022)]%
        {bayer2022role}
\bibfield{author}{\bibinfo{person}{Sarah Bayer}, \bibinfo{person}{Henner Gimpel}, {and} \bibinfo{person}{Moritz Markgraf}.} \bibinfo{year}{2022}\natexlab{}.
\newblock \showarticletitle{The role of domain expertise in trusting and following explainable AI decision support systems}.
\newblock \bibinfo{journal}{\emph{Journal of Decision Systems}} \bibinfo{volume}{32}, \bibinfo{number}{1} (\bibinfo{year}{2022}), \bibinfo{pages}{110--138}.
\newblock


\bibitem[Bertrand et~al\mbox{.}(2023)]%
        {bertrand2023questioning}
\bibfield{author}{\bibinfo{person}{Astrid Bertrand}, \bibinfo{person}{James~R Eagan}, {and} \bibinfo{person}{Winston Maxwell}.} \bibinfo{year}{2023}\natexlab{}.
\newblock \showarticletitle{Questioning the ability of feature-based explanations to empower non-experts in robo-advised financial decision-making}. In \bibinfo{booktitle}{\emph{Proceedings of the 2023 ACM Conference on Fairness, Accountability, and Transparency}}. \bibinfo{pages}{943--958}.
\newblock


\bibitem[Braun and Clarke(2006)]%
        {braun2006using}
\bibfield{author}{\bibinfo{person}{Virginia Braun} {and} \bibinfo{person}{Victoria Clarke}.} \bibinfo{year}{2006}\natexlab{}.
\newblock \showarticletitle{Using thematic analysis in psychology}.
\newblock \bibinfo{journal}{\emph{Qualitative research in psychology}} \bibinfo{volume}{3}, \bibinfo{number}{2} (\bibinfo{year}{2006}), \bibinfo{pages}{77--101}.
\newblock


\bibitem[Bu\c{c}inca et~al\mbox{.}(2021)]%
        {zana}
\bibfield{author}{\bibinfo{person}{Zana Bu\c{c}inca}, \bibinfo{person}{Maja~Barbara Malaya}, {and} \bibinfo{person}{Krzysztof~Z. Gajos}.} \bibinfo{year}{2021}\natexlab{}.
\newblock \showarticletitle{To Trust or to Think: Cognitive Forcing Functions Can Reduce Overreliance on AI in AI-Assisted Decision-Making}.
\newblock \bibinfo{journal}{\emph{Proc. ACM Hum.-Comput. Interact.}} \bibinfo{volume}{5}, \bibinfo{number}{CSCW1}, Article \bibinfo{articleno}{188} (\bibinfo{date}{apr} \bibinfo{year}{2021}), \bibinfo{numpages}{21}~pages.
\newblock
\urldef\tempurl%
\url{https://doi.org/10.1145/3449287}
\showDOI{\tempurl}


\bibitem[Bullock(2019)]%
        {bullock2019artificial}
\bibfield{author}{\bibinfo{person}{Justin~B Bullock}.} \bibinfo{year}{2019}\natexlab{}.
\newblock \showarticletitle{Artificial intelligence, discretion, and bureaucracy}.
\newblock \bibinfo{journal}{\emph{The American Review of Public Administration}} \bibinfo{volume}{49}, \bibinfo{number}{7} (\bibinfo{year}{2019}), \bibinfo{pages}{751--761}.
\newblock


\bibitem[Burnett(2020)]%
        {burnett2020explaining}
\bibfield{author}{\bibinfo{person}{Margaret Burnett}.} \bibinfo{year}{2020}\natexlab{}.
\newblock \showarticletitle{Explaining AI: fairly? well?}. In \bibinfo{booktitle}{\emph{Proceedings of the 25th International Conference on Intelligent User Interfaces}}. \bibinfo{pages}{1--2}.
\newblock


\bibitem[Chen and Barnes(2014)]%
        {chen2014human}
\bibfield{author}{\bibinfo{person}{Jessie~YC Chen} {and} \bibinfo{person}{Michael~J Barnes}.} \bibinfo{year}{2014}\natexlab{}.
\newblock \showarticletitle{Human--agent teaming for multirobot control: A review of human factors issues}.
\newblock \bibinfo{journal}{\emph{IEEE Transactions on Human-Machine Systems}} \bibinfo{volume}{44}, \bibinfo{number}{1} (\bibinfo{year}{2014}), \bibinfo{pages}{13--29}.
\newblock


\bibitem[Chromik et~al\mbox{.}(2021)]%
        {chromik}
\bibfield{author}{\bibinfo{person}{Michael Chromik}, \bibinfo{person}{Malin Eiband}, \bibinfo{person}{Felicitas Buchner}, \bibinfo{person}{Adrian Kr\"{u}ger}, {and} \bibinfo{person}{Andreas Butz}.} \bibinfo{year}{2021}\natexlab{}.
\newblock \showarticletitle{I Think I Get Your Point, AI! The Illusion of Explanatory Depth in Explainable AI}. In \bibinfo{booktitle}{\emph{Proceedings of the 26th International Conference on Intelligent User Interfaces}} (College Station, TX, USA) \emph{(\bibinfo{series}{IUI '21})}. \bibinfo{publisher}{Association for Computing Machinery}, \bibinfo{address}{New York, NY, USA}, \bibinfo{pages}{307–317}.
\newblock
\showISBNx{9781450380171}
\urldef\tempurl%
\url{https://doi.org/10.1145/3397481.3450644}
\showDOI{\tempurl}


\bibitem[Commission(2021)]%
        {com2021laying}
\bibfield{author}{\bibinfo{person}{EU Commission}.} \bibinfo{year}{2021}\natexlab{}.
\newblock \bibinfo{title}{Laying down harmonised rules on artificial intelligence (artificial intelligence act) and amending certain union legislative acts. Proposal for a regulation of the European parliament and of the council}.
\newblock
\newblock


\bibitem[Danks(2019)]%
        {danks2019value}
\bibfield{author}{\bibinfo{person}{David Danks}.} \bibinfo{year}{2019}\natexlab{}.
\newblock \showarticletitle{The value of trustworthy AI}. In \bibinfo{booktitle}{\emph{Proceedings of the 2019 AAAI/ACM Conference on AI, Ethics, and Society}}. \bibinfo{pages}{521--522}.
\newblock


\bibitem[de~Bruijn et~al\mbox{.}(2022)]%
        {de2022perils}
\bibfield{author}{\bibinfo{person}{Hans de Bruijn}, \bibinfo{person}{Martijn Warnier}, {and} \bibinfo{person}{Marijn Janssen}.} \bibinfo{year}{2022}\natexlab{}.
\newblock \showarticletitle{The perils and pitfalls of explainable AI: Strategies for explaining algorithmic decision-making}.
\newblock \bibinfo{journal}{\emph{Government information quarterly}} \bibinfo{volume}{39}, \bibinfo{number}{2} (\bibinfo{year}{2022}), \bibinfo{pages}{101666}.
\newblock


\bibitem[De~Visser et~al\mbox{.}(2020)]%
        {de2020towards}
\bibfield{author}{\bibinfo{person}{Ewart~J De~Visser}, \bibinfo{person}{Marieke~MM Peeters}, \bibinfo{person}{Malte~F Jung}, \bibinfo{person}{Spencer Kohn}, \bibinfo{person}{Tyler~H Shaw}, \bibinfo{person}{Richard Pak}, {and} \bibinfo{person}{Mark~A Neerincx}.} \bibinfo{year}{2020}\natexlab{}.
\newblock \showarticletitle{Towards a theory of longitudinal trust calibration in human--robot teams}.
\newblock \bibinfo{journal}{\emph{International journal of social robotics}} \bibinfo{volume}{12}, \bibinfo{number}{2} (\bibinfo{year}{2020}), \bibinfo{pages}{459--478}.
\newblock


\bibitem[Degachi et~al\mbox{.}(2024)]%
        {10.1145/3613905.3650825}
\bibfield{author}{\bibinfo{person}{Chadha Degachi}, \bibinfo{person}{Siddharth Mehrotra}, \bibinfo{person}{Mireia Yurrita}, \bibinfo{person}{Evangelos Niforatos}, {and} \bibinfo{person}{Myrthe~Lotte Tielman}.} \bibinfo{year}{2024}\natexlab{}.
\newblock \showarticletitle{Practising Appropriate Trust in Human-Centred AI Design}. In \bibinfo{booktitle}{\emph{Extended Abstracts of the CHI Conference on Human Factors in Computing Systems}} (Honolulu, HI, USA) \emph{(\bibinfo{series}{CHI EA '24})}. \bibinfo{publisher}{Association for Computing Machinery}, \bibinfo{address}{New York, NY, USA}, Article \bibinfo{articleno}{269}, \bibinfo{numpages}{8}~pages.
\newblock
\showISBNx{9798400703317}
\urldef\tempurl%
\url{https://doi.org/10.1145/3613905.3650825}
\showDOI{\tempurl}


\bibitem[Dodge et~al\mbox{.}(2019)]%
        {dodge2019}
\bibfield{author}{\bibinfo{person}{Jonathan Dodge}, \bibinfo{person}{Q~Vera Liao}, \bibinfo{person}{Yunfeng Zhang}, \bibinfo{person}{Rachel~KE Bellamy}, {and} \bibinfo{person}{Casey Dugan}.} \bibinfo{year}{2019}\natexlab{}.
\newblock \showarticletitle{Explaining models: an empirical study of how explanations impact fairness judgment}. In \bibinfo{booktitle}{\emph{Proceedings of the 24th international conference on intelligent user interfaces}}. \bibinfo{pages}{275--285}.
\newblock


\bibitem[Doshi-Velez and Kim(2017)]%
        {doshi2017towards}
\bibfield{author}{\bibinfo{person}{Finale Doshi-Velez} {and} \bibinfo{person}{Been Kim}.} \bibinfo{year}{2017}\natexlab{}.
\newblock \showarticletitle{Towards a rigorous science of interpretable machine learning}.
\newblock \bibinfo{journal}{\emph{arXiv preprint arXiv:1702.08608}} (\bibinfo{year}{2017}).
\newblock


\bibitem[Eiband et~al\mbox{.}(2019)]%
        {10.1145/3290607.3312787}
\bibfield{author}{\bibinfo{person}{Malin Eiband}, \bibinfo{person}{Daniel Buschek}, \bibinfo{person}{Alexander Kremer}, {and} \bibinfo{person}{Heinrich Hussmann}.} \bibinfo{year}{2019}\natexlab{}.
\newblock \showarticletitle{The Impact of Placebic Explanations on Trust in Intelligent Systems}. In \bibinfo{booktitle}{\emph{Extended Abstracts of the 2019 CHI Conference on Human Factors in Computing Systems}} (Glasgow, Scotland Uk) \emph{(\bibinfo{series}{CHI EA '19})}. \bibinfo{publisher}{Association for Computing Machinery}, \bibinfo{address}{New York, NY, USA}, \bibinfo{pages}{1–6}.
\newblock
\showISBNx{9781450359719}
\urldef\tempurl%
\url{https://doi.org/10.1145/3290607.3312787}
\showDOI{\tempurl}


\bibitem[Endsley(2023)]%
        {endsley2023ironies}
\bibfield{author}{\bibinfo{person}{Mica~R Endsley}.} \bibinfo{year}{2023}\natexlab{}.
\newblock \showarticletitle{Ironies of artificial intelligence}.
\newblock \bibinfo{journal}{\emph{Ergonomics}} \bibinfo{volume}{66}, \bibinfo{number}{11} (\bibinfo{year}{2023}), \bibinfo{pages}{1656--1668}.
\newblock


\bibitem[Ensign et~al\mbox{.}(2018)]%
        {ensign2018runaway}
\bibfield{author}{\bibinfo{person}{Danielle Ensign}, \bibinfo{person}{Sorelle~A Friedler}, \bibinfo{person}{Scott Neville}, \bibinfo{person}{Carlos Scheidegger}, {and} \bibinfo{person}{Suresh Venkatasubramanian}.} \bibinfo{year}{2018}\natexlab{}.
\newblock \showarticletitle{Runaway feedback loops in predictive policing}. In \bibinfo{booktitle}{\emph{Conference on fairness, accountability and transparency}}. PMLR, \bibinfo{pages}{160--171}.
\newblock


\bibitem[Faul et~al\mbox{.}(2009)]%
        {faul2009statistical}
\bibfield{author}{\bibinfo{person}{Franz Faul}, \bibinfo{person}{Edgar Erdfelder}, \bibinfo{person}{Axel Buchner}, {and} \bibinfo{person}{Albert-Georg Lang}.} \bibinfo{year}{2009}\natexlab{}.
\newblock \showarticletitle{Statistical power analyses using G* Power 3.1: Tests for correlation and regression analyses}.
\newblock \bibinfo{journal}{\emph{Behavior research methods}} \bibinfo{volume}{41}, \bibinfo{number}{4} (\bibinfo{year}{2009}), \bibinfo{pages}{1149--1160}.
\newblock


\bibitem[Ferguson(2018)]%
        {ferguson2018legal}
\bibfield{author}{\bibinfo{person}{Andrew~Guthrie Ferguson}.} \bibinfo{year}{2018}\natexlab{}.
\newblock \showarticletitle{The legal risks of big data policing}.
\newblock \bibinfo{journal}{\emph{Crim. Just.}}  \bibinfo{volume}{33} (\bibinfo{year}{2018}), \bibinfo{pages}{4}.
\newblock


\bibitem[Ferreira Gomes Centeio~Jorge et~al\mbox{.}(2021)]%
        {carolina}
\bibfield{author}{\bibinfo{person}{Carolina Ferreira Gomes Centeio~Jorge}, \bibinfo{person}{Siddharth Mehrotra}, \bibinfo{person}{Myrthe~L. Tielman}, {and} \bibinfo{person}{Catholijn~M. Jonker}.} \bibinfo{year}{2021}\natexlab{}.
\newblock \showarticletitle{Trust should correspond to Trustworthiness: a Formalization of Appropriate Mutual Trust in Human-Agent Teams}.
\newblock \bibinfo{journal}{\emph{Proceedings of the 22nd International Workshop on Trust in Agent Societies, London, UK}} (\bibinfo{year}{2021}).
\newblock


\bibitem[Gerber and Jackson(2017)]%
        {gerber2017justifying}
\bibfield{author}{\bibinfo{person}{Monica~M Gerber} {and} \bibinfo{person}{Jonathan Jackson}.} \bibinfo{year}{2017}\natexlab{}.
\newblock \showarticletitle{Justifying violence: legitimacy, ideology and public support for police use of force}.
\newblock \bibinfo{journal}{\emph{Psychology, crime \& law}} \bibinfo{volume}{23}, \bibinfo{number}{1} (\bibinfo{year}{2017}), \bibinfo{pages}{79--95}.
\newblock


\bibitem[Gerstner(2018)]%
        {gerstner2018predictive}
\bibfield{author}{\bibinfo{person}{Dominik Gerstner}.} \bibinfo{year}{2018}\natexlab{}.
\newblock \showarticletitle{Predictive policing in the context of residential burglary: An empirical illustration on the basis of a pilot project in Baden-W{\"u}rttemberg, Germany}.
\newblock \bibinfo{journal}{\emph{European Journal for Security Research}} \bibinfo{volume}{3}, \bibinfo{number}{2} (\bibinfo{year}{2018}), \bibinfo{pages}{115--138}.
\newblock


\bibitem[Gunning et~al\mbox{.}(2021)]%
        {gunning}
\bibfield{author}{\bibinfo{person}{David Gunning}, \bibinfo{person}{Eric Vorm}, \bibinfo{person}{Jennifer~Yunyan Wang}, {and} \bibinfo{person}{Matt Turek}.} \bibinfo{year}{2021}\natexlab{}.
\newblock \showarticletitle{DARPA's explainable AI (XAI) program: A retrospective}.
\newblock \bibinfo{journal}{\emph{Applied AI Letters}} \bibinfo{volume}{2}, \bibinfo{number}{4} (\bibinfo{year}{2021}), \bibinfo{pages}{e61}.
\newblock
\urldef\tempurl%
\url{https://doi.org/10.1002/ail2.61}
\showDOI{\tempurl}


\bibitem[Hardyns and Rummens(2018)]%
        {hardyns2018predictive}
\bibfield{author}{\bibinfo{person}{Wim Hardyns} {and} \bibinfo{person}{Anneleen Rummens}.} \bibinfo{year}{2018}\natexlab{}.
\newblock \showarticletitle{Predictive policing as a new tool for law enforcement? Recent developments and challenges}.
\newblock \bibinfo{journal}{\emph{European journal on criminal policy and research}}  \bibinfo{volume}{24} (\bibinfo{year}{2018}), \bibinfo{pages}{201--218}.
\newblock


\bibitem[He et~al\mbox{.}(2022)]%
        {he2022like}
\bibfield{author}{\bibinfo{person}{Gaole He}, \bibinfo{person}{Agathe Balayn}, \bibinfo{person}{Stefan Buijsman}, \bibinfo{person}{Jie Yang}, {and} \bibinfo{person}{Ujwal Gadiraju}.} \bibinfo{year}{2022}\natexlab{}.
\newblock \showarticletitle{It Is Like Finding a Polar Bear in the Savannah! Concept-Level AI Explanations with Analogical Inference from Commonsense Knowledge}. In \bibinfo{booktitle}{\emph{Proceedings of the AAAI Conference on Human Computation and Crowdsourcing}}, Vol.~\bibinfo{volume}{10}. \bibinfo{pages}{89--101}.
\newblock


\bibitem[Hohman et~al\mbox{.}(2019)]%
        {hohman2019telegam}
\bibfield{author}{\bibinfo{person}{Fred Hohman}, \bibinfo{person}{Arjun Srinivasan}, {and} \bibinfo{person}{Steven~M Drucker}.} \bibinfo{year}{2019}\natexlab{}.
\newblock \showarticletitle{TeleGam: Combining visualization and verbalization for interpretable machine learning}. In \bibinfo{booktitle}{\emph{2019 IEEE Visualization Conference (VIS)}}. IEEE, \bibinfo{pages}{151--155}.
\newblock


\bibitem[Jacovi et~al\mbox{.}(2021)]%
        {jacovi}
\bibfield{author}{\bibinfo{person}{Alon Jacovi}, \bibinfo{person}{Ana Marasovi{\'c}}, \bibinfo{person}{Tim Miller}, {and} \bibinfo{person}{Yoav Goldberg}.} \bibinfo{year}{2021}\natexlab{}.
\newblock \showarticletitle{Formalizing trust in artificial intelligence: Prerequisites, causes and goals of human trust in AI}. In \bibinfo{booktitle}{\emph{Proceedings of the 2021 ACM conference on fairness, accountability, and transparency}}. \bibinfo{pages}{624--635}.
\newblock


\bibitem[Jameson et~al\mbox{.}(2014)]%
        {jameson2014choice}
\bibfield{author}{\bibinfo{person}{Anthony Jameson}, \bibinfo{person}{Bettina Berendt}, \bibinfo{person}{Silvia Gabrielli}, \bibinfo{person}{Federica Cena}, \bibinfo{person}{Cristina Gena}, \bibinfo{person}{Fabiana Vernero}, \bibinfo{person}{Katharina Reinecke}, {et~al\mbox{.}}} \bibinfo{year}{2014}\natexlab{}.
\newblock \showarticletitle{Choice architecture for human-computer interaction}.
\newblock \bibinfo{journal}{\emph{Foundations and Trends{\textregistered} in Human--Computer Interaction}} \bibinfo{volume}{7}, \bibinfo{number}{1--2} (\bibinfo{year}{2014}), \bibinfo{pages}{1--235}.
\newblock


\bibitem[Kapania et~al\mbox{.}(2022)]%
        {shivani}
\bibfield{author}{\bibinfo{person}{Shivani Kapania}, \bibinfo{person}{Oliver Siy}, \bibinfo{person}{Gabe Clapper}, \bibinfo{person}{Azhagu~Meena SP}, {and} \bibinfo{person}{Nithya Sambasivan}.} \bibinfo{year}{2022}\natexlab{}.
\newblock \showarticletitle{”Because AI is 100\% Right and Safe”: User Attitudes and Sources of AI Authority in India}. In \bibinfo{booktitle}{\emph{Proceedings of the 2022 CHI Conference on Human Factors in Computing Systems}} \emph{(\bibinfo{series}{CHI '22})}. \bibinfo{publisher}{Association for Computing Machinery}, \bibinfo{address}{New York, NY, USA}, Article \bibinfo{articleno}{158}, \bibinfo{numpages}{18}~pages.
\newblock
\showISBNx{9781450391573}
\urldef\tempurl%
\url{https://doi.org/10.1145/3491102.3517533}
\showDOI{\tempurl}


\bibitem[Kaur et~al\mbox{.}(2020)]%
        {kaur2020interpreting}
\bibfield{author}{\bibinfo{person}{Harmanpreet Kaur}, \bibinfo{person}{Harsha Nori}, \bibinfo{person}{Samuel Jenkins}, \bibinfo{person}{Rich Caruana}, \bibinfo{person}{Hanna Wallach}, {and} \bibinfo{person}{Jennifer Wortman~Vaughan}.} \bibinfo{year}{2020}\natexlab{}.
\newblock \showarticletitle{Interpreting interpretability: understanding data scientists' use of interpretability tools for machine learning}. In \bibinfo{booktitle}{\emph{Proceedings of the 2020 CHI conference on human factors in computing systems}}. \bibinfo{pages}{1--14}.
\newblock


\bibitem[Khasawneh et~al\mbox{.}(2003)]%
        {khasawneh2003model}
\bibfield{author}{\bibinfo{person}{Mohammad~T Khasawneh}, \bibinfo{person}{Shannon~R Bowling}, \bibinfo{person}{Xiaochun Jiang}, \bibinfo{person}{Anand~K Gramopadhye}, {and} \bibinfo{person}{Brian~J Melloy}.} \bibinfo{year}{2003}\natexlab{}.
\newblock \showarticletitle{A model for predicting human trust in automated systems}.
\newblock \bibinfo{journal}{\emph{Origins}}  \bibinfo{volume}{5} (\bibinfo{year}{2003}).
\newblock


\bibitem[Kim et~al\mbox{.}(2015)]%
        {kim2015inferring}
\bibfield{author}{\bibinfo{person}{Been Kim}, \bibinfo{person}{Caleb~M Chacha}, {and} \bibinfo{person}{Julie~A Shah}.} \bibinfo{year}{2015}\natexlab{}.
\newblock \showarticletitle{Inferring team task plans from human meetings: A generative modeling approach with logic-based prior}.
\newblock \bibinfo{journal}{\emph{Journal of Artificial Intelligence Research}}  \bibinfo{volume}{52} (\bibinfo{year}{2015}), \bibinfo{pages}{361--398}.
\newblock


\bibitem[Kim et~al\mbox{.}(2018)]%
        {kim2018textual}
\bibfield{author}{\bibinfo{person}{Jinkyu Kim}, \bibinfo{person}{Anna Rohrbach}, \bibinfo{person}{Trevor Darrell}, \bibinfo{person}{John Canny}, {and} \bibinfo{person}{Zeynep Akata}.} \bibinfo{year}{2018}\natexlab{}.
\newblock \showarticletitle{Textual explanations for self-driving vehicles}. In \bibinfo{booktitle}{\emph{Proceedings of the European conference on computer vision (ECCV)}}. \bibinfo{pages}{563--578}.
\newblock


\bibitem[Lai and Tan(2019)]%
        {laiv}
\bibfield{author}{\bibinfo{person}{Vivian Lai} {and} \bibinfo{person}{Chenhao Tan}.} \bibinfo{year}{2019}\natexlab{}.
\newblock \showarticletitle{On Human Predictions with Explanations and Predictions of Machine Learning Models: A Case Study on Deception Detection}. In \bibinfo{booktitle}{\emph{Proceedings of the Conference on Fairness, Accountability, and Transparency}} (Atlanta, GA, USA) \emph{(\bibinfo{series}{FAT* '19})}. \bibinfo{publisher}{Association for Computing Machinery}, \bibinfo{address}{New York, NY, USA}, \bibinfo{pages}{29–38}.
\newblock
\showISBNx{9781450361255}
\urldef\tempurl%
\url{https://doi.org/10.1145/3287560.3287590}
\showDOI{\tempurl}


\bibitem[Lakkaraju et~al\mbox{.}(2016)]%
        {10.1145/2939672.2939874}
\bibfield{author}{\bibinfo{person}{Himabindu Lakkaraju}, \bibinfo{person}{Stephen~H. Bach}, {and} \bibinfo{person}{Jure Leskovec}.} \bibinfo{year}{2016}\natexlab{}.
\newblock \showarticletitle{Interpretable Decision Sets: A Joint Framework for Description and Prediction}. In \bibinfo{booktitle}{\emph{Proceedings of the 22nd ACM SIGKDD International Conference on Knowledge Discovery and Data Mining}} (San Francisco, California, USA) \emph{(\bibinfo{series}{KDD '16})}. \bibinfo{publisher}{Association for Computing Machinery}, \bibinfo{address}{New York, NY, USA}, \bibinfo{pages}{1675–1684}.
\newblock
\showISBNx{9781450342322}
\urldef\tempurl%
\url{https://doi.org/10.1145/2939672.2939874}
\showDOI{\tempurl}


\bibitem[Larasati et~al\mbox{.}(2023a)]%
        {larasati2023meaningful}
\bibfield{author}{\bibinfo{person}{Retno Larasati}, \bibinfo{person}{Anna De~Liddo}, {and} \bibinfo{person}{Enrico Motta}.} \bibinfo{year}{2023}\natexlab{a}.
\newblock \showarticletitle{Meaningful Explanation Effect on User’s Trust in an AI Medical System: Designing Explanations for Non-Expert Users}.
\newblock \bibinfo{journal}{\emph{ACM Transactions on Interactive Intelligent Systems}} \bibinfo{volume}{13}, \bibinfo{number}{4} (\bibinfo{year}{2023}), \bibinfo{pages}{1--39}.
\newblock


\bibitem[Larasati et~al\mbox{.}(2023b)]%
        {10.1145/3631614}
\bibfield{author}{\bibinfo{person}{Retno Larasati}, \bibinfo{person}{Anna De~Liddo}, {and} \bibinfo{person}{Enrico Motta}.} \bibinfo{year}{2023}\natexlab{b}.
\newblock \showarticletitle{Meaningful Explanation Effect on User’s Trust in an AI Medical System: Designing Explanations for Non-Expert Users}.
\newblock \bibinfo{journal}{\emph{ACM Trans. Interact. Intell. Syst.}} \bibinfo{volume}{13}, \bibinfo{number}{4}, Article \bibinfo{articleno}{30} (\bibinfo{date}{dec} \bibinfo{year}{2023}), \bibinfo{numpages}{39}~pages.
\newblock
\showISSN{2160-6455}
\urldef\tempurl%
\url{https://doi.org/10.1145/3631614}
\showDOI{\tempurl}


\bibitem[Lee and See(2004)]%
        {leesee}
\bibfield{author}{\bibinfo{person}{John~D Lee} {and} \bibinfo{person}{Katrina~A See}.} \bibinfo{year}{2004}\natexlab{}.
\newblock \showarticletitle{Trust in automation: Designing for appropriate reliance}.
\newblock \bibinfo{journal}{\emph{Human factors}} \bibinfo{volume}{46}, \bibinfo{number}{1} (\bibinfo{year}{2004}), \bibinfo{pages}{50--80}.
\newblock


\bibitem[Lee and Chew(2023)]%
        {10.1145/3610218}
\bibfield{author}{\bibinfo{person}{Min~Hun Lee} {and} \bibinfo{person}{Chong~Jun Chew}.} \bibinfo{year}{2023}\natexlab{}.
\newblock \showarticletitle{Understanding the Effect of Counterfactual Explanations on Trust and Reliance on AI for Human-AI Collaborative Clinical Decision Making}.
\newblock \bibinfo{journal}{\emph{Proc. ACM Hum.-Comput. Interact.}} \bibinfo{volume}{7}, \bibinfo{number}{CSCW2}, Article \bibinfo{articleno}{369} (\bibinfo{date}{oct} \bibinfo{year}{2023}), \bibinfo{numpages}{22}~pages.
\newblock
\urldef\tempurl%
\url{https://doi.org/10.1145/3610218}
\showDOI{\tempurl}


\bibitem[Leese(2024)]%
        {leese2024staying}
\bibfield{author}{\bibinfo{person}{Matthias Leese}.} \bibinfo{year}{2024}\natexlab{}.
\newblock \showarticletitle{Staying in control of technology: Predictive policing, democracy, and digital sovereignty}.
\newblock \bibinfo{journal}{\emph{Democratization}} \bibinfo{volume}{31}, \bibinfo{number}{5} (\bibinfo{year}{2024}), \bibinfo{pages}{963--978}.
\newblock


\bibitem[Liao and Sundar(2022)]%
        {liao2022designing}
\bibfield{author}{\bibinfo{person}{Q.Vera Liao} {and} \bibinfo{person}{S.~Shyam Sundar}.} \bibinfo{year}{2022}\natexlab{}.
\newblock \showarticletitle{Designing for Responsible Trust in AI Systems: A Communication Perspective}. In \bibinfo{booktitle}{\emph{2022 ACM Conference on Fairness, Accountability, and Transparency}} (Seoul, Republic of Korea) \emph{(\bibinfo{series}{FAccT '22})}. \bibinfo{publisher}{Association for Computing Machinery}, \bibinfo{address}{New York, NY, USA}, \bibinfo{pages}{1257–1268}.
\newblock
\showISBNx{9781450393522}
\urldef\tempurl%
\url{https://doi.org/10.1145/3531146.3533182}
\showDOI{\tempurl}


\bibitem[Lum and Isaac(2016)]%
        {lum2016predict}
\bibfield{author}{\bibinfo{person}{Kristian Lum} {and} \bibinfo{person}{William Isaac}.} \bibinfo{year}{2016}\natexlab{}.
\newblock \showarticletitle{To predict and serve?}
\newblock \bibinfo{journal}{\emph{Significance}} \bibinfo{volume}{13}, \bibinfo{number}{5} (\bibinfo{year}{2016}), \bibinfo{pages}{14--19}.
\newblock


\bibitem[Lyons et~al\mbox{.}(2022)]%
        {lyons2022}
\bibfield{author}{\bibinfo{person}{Henrietta Lyons}, \bibinfo{person}{Senuri Wijenayake}, \bibinfo{person}{Tim Miller}, {and} \bibinfo{person}{Eduardo Velloso}.} \bibinfo{year}{2022}\natexlab{}.
\newblock \showarticletitle{What’s the Appeal? Perceptions of Review Processes for Algorithmic Decisions}. In \bibinfo{booktitle}{\emph{Proceedings of the 2022 CHI Conference on Human Factors in Computing Systems}} (<conf-loc>, <city>New Orleans</city>, <state>LA</state>, <country>USA</country>, </conf-loc>) \emph{(\bibinfo{series}{CHI '22})}. \bibinfo{publisher}{Association for Computing Machinery}, \bibinfo{address}{New York, NY, USA}, Article \bibinfo{articleno}{580}, \bibinfo{numpages}{15}~pages.
\newblock
\showISBNx{9781450391573}
\urldef\tempurl%
\url{https://doi.org/10.1145/3491102.3517606}
\showDOI{\tempurl}


\bibitem[Marda and Narayan(2020)]%
        {marda2020data}
\bibfield{author}{\bibinfo{person}{Vidushi Marda} {and} \bibinfo{person}{Shivangi Narayan}.} \bibinfo{year}{2020}\natexlab{}.
\newblock \showarticletitle{Data in New Delhi's predictive policing system}. In \bibinfo{booktitle}{\emph{Proceedings of the 2020 conference on fairness, accountability, and transparency}}. \bibinfo{pages}{317--324}.
\newblock


\bibitem[Max(2022)]%
        {max2022soundthinking}
\bibfield{author}{\bibinfo{person}{Brendan Max}.} \bibinfo{year}{2022}\natexlab{}.
\newblock \showarticletitle{SoundThinking's Black-Box Gunshot Detection Method: Untested and Unvetted Tech Flourishes in the Criminal Justice System}.
\newblock \bibinfo{journal}{\emph{Stan. Tech. L. Rev.}}  \bibinfo{volume}{26} (\bibinfo{year}{2022}), \bibinfo{pages}{193}.
\newblock


\bibitem[Mayer et~al\mbox{.}(1995)]%
        {mayer1995integrative}
\bibfield{author}{\bibinfo{person}{Roger~C Mayer}, \bibinfo{person}{James~H Davis}, {and} \bibinfo{person}{F~David Schoorman}.} \bibinfo{year}{1995}\natexlab{}.
\newblock \showarticletitle{An integrative model of organizational trust}.
\newblock \bibinfo{journal}{\emph{Academy of management review}} \bibinfo{volume}{20}, \bibinfo{number}{3} (\bibinfo{year}{1995}), \bibinfo{pages}{709--734}.
\newblock


\bibitem[McBride and Morgan(2010)]%
        {mcbride2010trust}
\bibfield{author}{\bibinfo{person}{Maranda McBride} {and} \bibinfo{person}{Shona Morgan}.} \bibinfo{year}{2010}\natexlab{}.
\newblock \showarticletitle{Trust calibration for automated decision aids}.
\newblock \bibinfo{journal}{\emph{Institute for Homeland Security Solutions}} (\bibinfo{year}{2010}), \bibinfo{pages}{1--11}.
\newblock


\bibitem[McDaniel and Pease(2021)]%
        {mcdaniel2021predictive}
\bibfield{author}{\bibinfo{person}{John McDaniel} {and} \bibinfo{person}{Ken Pease}.} \bibinfo{year}{2021}\natexlab{}.
\newblock \bibinfo{booktitle}{\emph{Predictive policing and artificial intelligence}}.
\newblock \bibinfo{publisher}{Routledge}.
\newblock


\bibitem[McGuirl and Sarter(2006)]%
        {mcguirl2006supporting}
\bibfield{author}{\bibinfo{person}{John~M McGuirl} {and} \bibinfo{person}{Nadine~B Sarter}.} \bibinfo{year}{2006}\natexlab{}.
\newblock \showarticletitle{Supporting trust calibration and the effective use of decision aids by presenting dynamic system confidence information}.
\newblock \bibinfo{journal}{\emph{Human factors}} \bibinfo{volume}{48}, \bibinfo{number}{4} (\bibinfo{year}{2006}), \bibinfo{pages}{656--665}.
\newblock


\bibitem[Mehrabi et~al\mbox{.}(2021)]%
        {mehrabi2021survey}
\bibfield{author}{\bibinfo{person}{Ninareh Mehrabi}, \bibinfo{person}{Fred Morstatter}, \bibinfo{person}{Nripsuta Saxena}, \bibinfo{person}{Kristina Lerman}, {and} \bibinfo{person}{Aram Galstyan}.} \bibinfo{year}{2021}\natexlab{}.
\newblock \showarticletitle{A survey on bias and fairness in machine learning}.
\newblock \bibinfo{journal}{\emph{ACM Computing Surveys (CSUR)}} \bibinfo{volume}{54}, \bibinfo{number}{6} (\bibinfo{year}{2021}), \bibinfo{pages}{1--35}.
\newblock


\bibitem[Mehrotra et~al\mbox{.}(2024a)]%
        {mehrotra2023systematic}
\bibfield{author}{\bibinfo{person}{Siddharth Mehrotra}, \bibinfo{person}{Chadha Degachi}, \bibinfo{person}{Oleksandra Vereschak}, \bibinfo{person}{Catholijn~M. Jonker}, {and} \bibinfo{person}{Myrthe~L. Tielman}.} \bibinfo{year}{2024}\natexlab{a}.
\newblock \showarticletitle{A Systematic Review on Fostering Appropriate Trust in Human-AI Interaction: Trends, Opportunities and Challenges}.
\newblock \bibinfo{journal}{\emph{ACM J. Responsib. Comput.}} \bibinfo{volume}{1}, \bibinfo{number}{4}, Article \bibinfo{articleno}{26} (\bibinfo{date}{Nov.} \bibinfo{year}{2024}), \bibinfo{numpages}{45}~pages.
\newblock
\urldef\tempurl%
\url{https://doi.org/10.1145/3696449}
\showDOI{\tempurl}


\bibitem[Mehrotra et~al\mbox{.}(2023)]%
        {mehrotra2023building}
\bibfield{author}{\bibinfo{person}{Siddharth Mehrotra}, \bibinfo{person}{Carolina~Centeio Jorge}, \bibinfo{person}{Catholijn~M Jonker}, {and} \bibinfo{person}{Myrthe~L Tielman}.} \bibinfo{year}{2023}\natexlab{}.
\newblock \showarticletitle{Building Appropriate Trust in AI: The Significance of Integrity-Centered Explanations.}. In \bibinfo{booktitle}{\emph{Volume 368: HHAI 2023: Augmenting Human Intellect}}. \bibinfo{pages}{436--439}.
\newblock
\urldef\tempurl%
\url{https://doi.org/10.3233/FAIA230121}
\showDOI{\tempurl}


\bibitem[Mehrotra et~al\mbox{.}(2024b)]%
        {mehrotraintegrity}
\bibfield{author}{\bibinfo{person}{Siddharth Mehrotra}, \bibinfo{person}{Carolina~Centeio Jorge}, \bibinfo{person}{Catholijn~M. Jonker}, {and} \bibinfo{person}{Myrthe~L. Tielman}.} \bibinfo{year}{2024}\natexlab{b}.
\newblock \showarticletitle{Integrity-based Explanations for Fostering Appropriate Trust in AI Agents}.
\newblock \bibinfo{journal}{\emph{ACM Trans. Interact. Intell. Syst.}} \bibinfo{volume}{14}, \bibinfo{number}{1}, Article \bibinfo{articleno}{4} (\bibinfo{date}{jan} \bibinfo{year}{2024}), \bibinfo{numpages}{36}~pages.
\newblock
\showISSN{2160-6455}
\urldef\tempurl%
\url{https://doi.org/10.1145/3610578}
\showDOI{\tempurl}


\bibitem[Meijer et~al\mbox{.}(2021)]%
        {meijer2021algorithmization}
\bibfield{author}{\bibinfo{person}{Albert Meijer}, \bibinfo{person}{Lukas Lorenz}, {and} \bibinfo{person}{Martijn Wessels}.} \bibinfo{year}{2021}\natexlab{}.
\newblock \showarticletitle{Algorithmization of bureaucratic organizations: Using a practice lens to study how context shapes predictive policing systems}.
\newblock \bibinfo{journal}{\emph{Public Administration Review}} \bibinfo{volume}{81}, \bibinfo{number}{5} (\bibinfo{year}{2021}), \bibinfo{pages}{837--846}.
\newblock


\bibitem[Meijer and Wessels(2019)]%
        {meijer2019predictive}
\bibfield{author}{\bibinfo{person}{Albert Meijer} {and} \bibinfo{person}{Martijn Wessels}.} \bibinfo{year}{2019}\natexlab{}.
\newblock \showarticletitle{Predictive policing: Review of benefits and drawbacks}.
\newblock \bibinfo{journal}{\emph{International Journal of Public Administration}} \bibinfo{volume}{42}, \bibinfo{number}{12} (\bibinfo{year}{2019}), \bibinfo{pages}{1031--1039}.
\newblock


\bibitem[Merritt(2011)]%
        {merritt2011affective}
\bibfield{author}{\bibinfo{person}{Stephanie~M Merritt}.} \bibinfo{year}{2011}\natexlab{}.
\newblock \showarticletitle{Affective processes in human--automation interactions}.
\newblock \bibinfo{journal}{\emph{Human Factors}} \bibinfo{volume}{53}, \bibinfo{number}{4} (\bibinfo{year}{2011}), \bibinfo{pages}{356--370}.
\newblock


\bibitem[Miller(2023)]%
        {miller2023explainable}
\bibfield{author}{\bibinfo{person}{Tim Miller}.} \bibinfo{year}{2023}\natexlab{}.
\newblock \showarticletitle{Explainable AI is Dead, Long Live Explainable AI! Hypothesis-driven Decision Support using Evaluative AI}. In \bibinfo{booktitle}{\emph{Proceedings of the 2023 ACM Conference on Fairness, Accountability, and Transparency}}. \bibinfo{pages}{333--342}.
\newblock


\bibitem[Mohseni et~al\mbox{.}(2018)]%
        {Mohseni2018ASO}
\bibfield{author}{\bibinfo{person}{Sina Mohseni}, \bibinfo{person}{Niloofar Zarei}, {and} \bibinfo{person}{Eric~D. Ragan}.} \bibinfo{year}{2018}\natexlab{}.
\newblock \showarticletitle{A Survey of Evaluation Methods and Measures for Interpretable Machine Learning}.
\newblock \bibinfo{journal}{\emph{ArXiv}}  \bibinfo{volume}{abs/1811.11839} (\bibinfo{year}{2018}).
\newblock
\urldef\tempurl%
\url{https://api.semanticscholar.org/CorpusID:54087635}
\showURL{%
\tempurl}


\bibitem[Mugari and Obioha(2021)]%
        {mugari2021predictive}
\bibfield{author}{\bibinfo{person}{Ishmael Mugari} {and} \bibinfo{person}{Emeka~E Obioha}.} \bibinfo{year}{2021}\natexlab{}.
\newblock \showarticletitle{Predictive policing and crime control in the United States of America and Europe: Trends in a decade of research and the future of predictive policing}.
\newblock \bibinfo{journal}{\emph{Social sciences}} \bibinfo{volume}{10}, \bibinfo{number}{6} (\bibinfo{year}{2021}), \bibinfo{pages}{234}.
\newblock


\bibitem[Nunes and Jannach(2017)]%
        {nunes2017systematic}
\bibfield{author}{\bibinfo{person}{Ingrid Nunes} {and} \bibinfo{person}{Dietmar Jannach}.} \bibinfo{year}{2017}\natexlab{}.
\newblock \showarticletitle{A systematic review and taxonomy of explanations in decision support and recommender systems}.
\newblock \bibinfo{journal}{\emph{User Modeling and User-Adapted Interaction}}  \bibinfo{volume}{27} (\bibinfo{year}{2017}), \bibinfo{pages}{393--444}.
\newblock


\bibitem[Okamura and Yamada(2020)]%
        {okamura2020adaptive}
\bibfield{author}{\bibinfo{person}{Kazuo Okamura} {and} \bibinfo{person}{Seiji Yamada}.} \bibinfo{year}{2020}\natexlab{}.
\newblock \showarticletitle{Adaptive trust calibration for human-AI collaboration}.
\newblock \bibinfo{journal}{\emph{PLoS ONE}} \bibinfo{volume}{15}, \bibinfo{number}{2} (\bibinfo{year}{2020}).
\newblock


\bibitem[Oosterloo et~al\mbox{.}(2018)]%
        {oosterloo2018politics}
\bibfield{author}{\bibinfo{person}{Serena Oosterloo}, \bibinfo{person}{Gerwin van Schie}, \bibinfo{person}{Jo Bates}, \bibinfo{person}{Paul Clough}, \bibinfo{person}{Robert J{\"a}schke}, \bibinfo{person}{Jahna Otterbacher}, {et~al\mbox{.}}} \bibinfo{year}{2018}\natexlab{}.
\newblock \showarticletitle{The politics and biases of the “crime anticipation system” of the Dutch police}. In \bibinfo{booktitle}{\emph{Proceedings of the international workshop on bias in information, algorithms, and systems}}, Vol.~\bibinfo{volume}{2103}. CEUR WS, \bibinfo{pages}{30--41}.
\newblock


\bibitem[Ososky et~al\mbox{.}(2013)]%
        {ososky2013building}
\bibfield{author}{\bibinfo{person}{Scott Ososky}, \bibinfo{person}{David Schuster}, \bibinfo{person}{Elizabeth Phillips}, {and} \bibinfo{person}{Florian~G Jentsch}.} \bibinfo{year}{2013}\natexlab{}.
\newblock \showarticletitle{Building appropriate trust in human-robot teams}. In \bibinfo{booktitle}{\emph{2013 AAAI spring symposium series}}.
\newblock


\bibitem[Parasuraman and Riley(1997)]%
        {parasuraman1997humans}
\bibfield{author}{\bibinfo{person}{Raja Parasuraman} {and} \bibinfo{person}{Victor Riley}.} \bibinfo{year}{1997}\natexlab{}.
\newblock \showarticletitle{Humans and automation: Use, misuse, disuse, abuse}.
\newblock \bibinfo{journal}{\emph{Human factors}} \bibinfo{volume}{39}, \bibinfo{number}{2} (\bibinfo{year}{1997}), \bibinfo{pages}{230--253}.
\newblock


\bibitem[Park et~al\mbox{.}(2018)]%
        {park2018multimodal}
\bibfield{author}{\bibinfo{person}{Dong~Huk Park}, \bibinfo{person}{Lisa~Anne Hendricks}, \bibinfo{person}{Zeynep Akata}, \bibinfo{person}{Anna Rohrbach}, \bibinfo{person}{Bernt Schiele}, \bibinfo{person}{Trevor Darrell}, {and} \bibinfo{person}{Marcus Rohrbach}.} \bibinfo{year}{2018}\natexlab{}.
\newblock \showarticletitle{Multimodal explanations: Justifying decisions and pointing to the evidence}. In \bibinfo{booktitle}{\emph{Proceedings of the IEEE conference on computer vision and pattern recognition}}. \bibinfo{pages}{8779--8788}.
\newblock


\bibitem[Ribera and Lapedriza~Garc{\'\i}a(2019)]%
        {ribera2019can}
\bibfield{author}{\bibinfo{person}{Mireia Ribera} {and} \bibinfo{person}{{\`A}gata Lapedriza~Garc{\'\i}a}.} \bibinfo{year}{2019}\natexlab{}.
\newblock \showarticletitle{Can we do better explanations? A proposal of user-centered explainable AI}. CEUR Workshop Proceedings.
\newblock


\bibitem[Robbemond et~al\mbox{.}(2022)]%
        {robbemond2022understanding}
\bibfield{author}{\bibinfo{person}{Vincent Robbemond}, \bibinfo{person}{Oana Inel}, {and} \bibinfo{person}{Ujwal Gadiraju}.} \bibinfo{year}{2022}\natexlab{}.
\newblock \showarticletitle{Understanding the Role of Explanation Modality in AI-assisted Decision-making}. In \bibinfo{booktitle}{\emph{Proceedings of the 30th ACM Conference on User Modeling, Adaptation and Personalization}}. \bibinfo{pages}{223--233}.
\newblock


\bibitem[Robinette et~al\mbox{.}(2016)]%
        {robinette2016overtrust}
\bibfield{author}{\bibinfo{person}{Paul Robinette}, \bibinfo{person}{Wenchen Li}, \bibinfo{person}{Robert Allen}, \bibinfo{person}{Ayanna~M Howard}, {and} \bibinfo{person}{Alan~R Wagner}.} \bibinfo{year}{2016}\natexlab{}.
\newblock \showarticletitle{Overtrust of robots in emergency evacuation scenarios}. In \bibinfo{booktitle}{\emph{2016 11th ACM/IEEE International Conference on Human-Robot Interaction (HRI)}}. IEEE, \bibinfo{pages}{101--108}.
\newblock


\bibitem[Schaffer et~al\mbox{.}(2019)]%
        {schaffer2019can}
\bibfield{author}{\bibinfo{person}{James Schaffer}, \bibinfo{person}{John O'Donovan}, \bibinfo{person}{James Michaelis}, \bibinfo{person}{Adrienne Raglin}, {and} \bibinfo{person}{Tobias H{\"o}llerer}.} \bibinfo{year}{2019}\natexlab{}.
\newblock \showarticletitle{I can do better than your AI: expertise and explanations}. In \bibinfo{booktitle}{\emph{Proceedings of the 24th International Conference on Intelligent User Interfaces}}. \bibinfo{pages}{240--251}.
\newblock


\bibitem[Schlicker and Langer(2021)]%
        {Schlicker}
\bibfield{author}{\bibinfo{person}{Nadine Schlicker} {and} \bibinfo{person}{Markus Langer}.} \bibinfo{year}{2021}\natexlab{}.
\newblock \showarticletitle{Towards Warranted Trust: A Model on the Relation Between Actual and Perceived System Trustworthiness}. In \bibinfo{booktitle}{\emph{Proceedings of Mensch Und Computer 2021}} (Ingolstadt, Germany) \emph{(\bibinfo{series}{MuC '21})}. \bibinfo{publisher}{Association for Computing Machinery}, \bibinfo{address}{New York, NY, USA}, \bibinfo{pages}{325–329}.
\newblock
\showISBNx{9781450386456}
\urldef\tempurl%
\url{https://doi.org/10.1145/3473856.3474018}
\showDOI{\tempurl}


\bibitem[Schoeffer et~al\mbox{.}(2022)]%
        {schoeffer2022there}
\bibfield{author}{\bibinfo{person}{Jakob Schoeffer}, \bibinfo{person}{Niklas Kuehl}, {and} \bibinfo{person}{Yvette Machowski}.} \bibinfo{year}{2022}\natexlab{}.
\newblock \showarticletitle{“There is not enough information”: On the effects of explanations on perceptions of informational fairness and trustworthiness in automated decision-making}. In \bibinfo{booktitle}{\emph{Proceedings of the 2022 ACM Conference on Fairness, Accountability, and Transparency}}. \bibinfo{pages}{1616--1628}.
\newblock


\bibitem[Selten et~al\mbox{.}(2023)]%
        {selten}
\bibfield{author}{\bibinfo{person}{Friso Selten}, \bibinfo{person}{Marcel Robeer}, {and} \bibinfo{person}{Stephan Grimmelikhuijsen}.} \bibinfo{year}{2023}\natexlab{}.
\newblock \showarticletitle{‘Just like I thought’: Street-level bureaucrats trust AI recommendations if they confirm their professional judgment}.
\newblock \bibinfo{journal}{\emph{Public Administration Review}} \bibinfo{volume}{83}, \bibinfo{number}{2} (\bibinfo{year}{2023}), \bibinfo{pages}{263--278}.
\newblock


\bibitem[Shively et~al\mbox{.}(2018)]%
        {shively2018human}
\bibfield{author}{\bibinfo{person}{R~Jay Shively}, \bibinfo{person}{Joel Lachter}, \bibinfo{person}{Summer~L Brandt}, \bibinfo{person}{Michael Matessa}, \bibinfo{person}{Vernol Battiste}, {and} \bibinfo{person}{Walter~W Johnson}.} \bibinfo{year}{2018}\natexlab{}.
\newblock \showarticletitle{Why human-autonomy teaming?}. In \bibinfo{booktitle}{\emph{Advances in Neuroergonomics and Cognitive Engineering: Proceedings of the AHFE 2017 International Conference on Neuroergonomics and Cognitive Engineering, July 17--21, 2017, The Westin Bonaventure Hotel, Los Angeles, California, USA 8}}. Springer, \bibinfo{pages}{3--11}.
\newblock


\bibitem[Simkute et~al\mbox{.}(2020)]%
        {simkute}
\bibfield{author}{\bibinfo{person}{Auste Simkute}, \bibinfo{person}{Ewa Luger}, \bibinfo{person}{Mike Evans}, {and} \bibinfo{person}{Rhianne Jones}.} \bibinfo{year}{2020}\natexlab{}.
\newblock \showarticletitle{Experts in the Shadow of Algorithmic Systems: Exploring Intelligibility in a Decision-Making Context}. In \bibinfo{booktitle}{\emph{Companion Publication of the 2020 ACM Designing Interactive Systems Conference}} (Eindhoven, Netherlands) \emph{(\bibinfo{series}{DIS' 20 Companion})}. \bibinfo{publisher}{Association for Computing Machinery}, \bibinfo{address}{New York, NY, USA}, \bibinfo{pages}{263–268}.
\newblock
\showISBNx{9781450379878}
\urldef\tempurl%
\url{https://doi.org/10.1145/3393914.3395862}
\showDOI{\tempurl}


\bibitem[Sperrle et~al\mbox{.}(2020)]%
        {sperrle2020should}
\bibfield{author}{\bibinfo{person}{Fabian Sperrle}, \bibinfo{person}{Mennatallah El-Assady}, \bibinfo{person}{Grace Guo}, \bibinfo{person}{Duen~Horng Chau}, \bibinfo{person}{Alex Endert}, {and} \bibinfo{person}{Daniel Keim}.} \bibinfo{year}{2020}\natexlab{}.
\newblock \showarticletitle{Should we trust (x) AI? Design dimensions for structured experimental evaluations}.
\newblock \bibinfo{journal}{\emph{arXiv preprint arXiv:2009.06433}} (\bibinfo{year}{2020}).
\newblock


\bibitem[Szymanski et~al\mbox{.}(2021)]%
        {szymanski2021visual}
\bibfield{author}{\bibinfo{person}{Maxwell Szymanski}, \bibinfo{person}{Martijn Millecamp}, {and} \bibinfo{person}{Katrien Verbert}.} \bibinfo{year}{2021}\natexlab{}.
\newblock \showarticletitle{Visual, textual or hybrid: the effect of user expertise on different explanations}. In \bibinfo{booktitle}{\emph{26th International Conference on Intelligent User Interfaces}}. \bibinfo{pages}{109--119}.
\newblock


\bibitem[Tolmeijer et~al\mbox{.}(2022)]%
        {suzanne2022capable}
\bibfield{author}{\bibinfo{person}{Suzanne Tolmeijer}, \bibinfo{person}{Markus Christen}, \bibinfo{person}{Serhiy Kandul}, \bibinfo{person}{Markus Kneer}, {and} \bibinfo{person}{Abraham Bernstein}.} \bibinfo{year}{2022}\natexlab{}.
\newblock \showarticletitle{Capable but Amoral? Comparing AI and Human Expert Collaboration in Ethical Decision Making}. In \bibinfo{booktitle}{\emph{CHI Conference on Human Factors in Computing Systems}}. \bibinfo{pages}{1--17}.
\newblock


\bibitem[Ulfert et~al\mbox{.}(2023)]%
        {ulfert2023shaping}
\bibfield{author}{\bibinfo{person}{Anna-Sophie Ulfert}, \bibinfo{person}{Eleni Georganta}, \bibinfo{person}{Carolina Centeio~Jorge}, \bibinfo{person}{Siddharth Mehrotra}, {and} \bibinfo{person}{Myrthe Tielman}.} \bibinfo{year}{2023}\natexlab{}.
\newblock \showarticletitle{Shaping a multidisciplinary understanding of team trust in human-AI teams: a theoretical framework}.
\newblock \bibinfo{journal}{\emph{European Journal of Work and Organizational Psychology}} (\bibinfo{year}{2023}), \bibinfo{pages}{1--14}.
\newblock


\bibitem[Vasconcelos et~al\mbox{.}(2023)]%
        {10.1145/3579605}
\bibfield{author}{\bibinfo{person}{Helena Vasconcelos}, \bibinfo{person}{Matthew J\"{o}rke}, \bibinfo{person}{Madeleine Grunde-McLaughlin}, \bibinfo{person}{Tobias Gerstenberg}, \bibinfo{person}{Michael~S. Bernstein}, {and} \bibinfo{person}{Ranjay Krishna}.} \bibinfo{year}{2023}\natexlab{}.
\newblock \showarticletitle{Explanations Can Reduce Overreliance on AI Systems During Decision-Making}.
\newblock \bibinfo{journal}{\emph{Proc. ACM Hum.-Comput. Interact.}} \bibinfo{volume}{7}, \bibinfo{number}{CSCW1}, Article \bibinfo{articleno}{129} (\bibinfo{date}{apr} \bibinfo{year}{2023}), \bibinfo{numpages}{38}~pages.
\newblock
\urldef\tempurl%
\url{https://doi.org/10.1145/3579605}
\showDOI{\tempurl}


\bibitem[Wang et~al\mbox{.}(2019)]%
        {wang2019designing}
\bibfield{author}{\bibinfo{person}{Danding Wang}, \bibinfo{person}{Qian Yang}, \bibinfo{person}{Ashraf Abdul}, {and} \bibinfo{person}{Brian~Y Lim}.} \bibinfo{year}{2019}\natexlab{}.
\newblock \showarticletitle{Designing theory-driven user-centric explainable AI}. In \bibinfo{booktitle}{\emph{Proceedings of the 2019 CHI conference on human factors in computing systems}}. \bibinfo{pages}{1--15}.
\newblock


\bibitem[Wischnewski et~al\mbox{.}(2023)]%
        {wischnewski2023measuring}
\bibfield{author}{\bibinfo{person}{Magdalena Wischnewski}, \bibinfo{person}{Nicole Kr{\"a}mer}, {and} \bibinfo{person}{Emmanuel M{\"u}ller}.} \bibinfo{year}{2023}\natexlab{}.
\newblock \showarticletitle{Measuring and Understanding Trust Calibrations for Automated Systems: A Survey of the State-Of-The-Art and Future Directions}. In \bibinfo{booktitle}{\emph{Proceedings of the 2023 CHI Conference on Human Factors in Computing Systems}}. \bibinfo{pages}{1--16}.
\newblock


\bibitem[Yang et~al\mbox{.}(2020)]%
        {yang2020visual}
\bibfield{author}{\bibinfo{person}{Fumeng Yang}, \bibinfo{person}{Zhuanyi Huang}, \bibinfo{person}{Jean Scholtz}, {and} \bibinfo{person}{Dustin~L Arendt}.} \bibinfo{year}{2020}\natexlab{}.
\newblock \showarticletitle{How do visual explanations foster end users' appropriate trust in machine learning?}. In \bibinfo{booktitle}{\emph{Proceedings of the 25th international conference on intelligent user interfaces}}. \bibinfo{pages}{189--201}.
\newblock


\bibitem[Zhang et~al\mbox{.}(2020)]%
        {zhang2020effect}
\bibfield{author}{\bibinfo{person}{Yunfeng Zhang}, \bibinfo{person}{Q~Vera Liao}, {and} \bibinfo{person}{Rachel~KE Bellamy}.} \bibinfo{year}{2020}\natexlab{}.
\newblock \showarticletitle{Effect of confidence and explanation on accuracy and trust calibration in AI-assisted decision making}. In \bibinfo{booktitle}{\emph{Proceedings of the 2020 Conference on Fairness, Accountability, and Transparency}}. \bibinfo{pages}{295--305}.
\newblock


\end{thebibliography}
\end{document}